\documentclass[12pt,preprint]{aastex}
\usepackage{graphicx}

\shorttitle{
Evolution of Massive Stars \\
 Up to the End of Central Oxygen Burning}
\shortauthors{Mounib et al.}

\def\msun{\, {{\rm M_{\odot}}}}       

\def\beqa{\begingroup{\global\advance\eqnuma by 1}$$}   

\def\chap#1{ \global \advance\chapnum by 1  \chpreset \chskipt
             \line{ \the\chapnum . \chpindent #1\hfill}
             \penalty 100000 \chskipl \penalty 100000 }
\def\:{\mskip\medmuskip}                         
 \def\sqr#1#2{{\vcenter{\hrule height.#2pt
     \hbox{\vrule width.#2pt height#1pt \kern#1pt
           \vrule width.#2pt}
           \hrule height.#2pt}}}
 
\begin{document}
\title{
Evolution of Massive Stars \\
 Up to the End of Central Oxygen Burning}

\author{M. F. El Eid\altaffilmark{1}}
\affil{Department of Physics, \\
American University of Beirut (AUB), 
Beirut, Lebanon}
\email{meid@aub.edu.lb}
\and
\author{B. S. Meyer and L.-S. The}
\affil{Department of Physics and Astronomy, , Kinard Laboratory of
Physics, \\
Clemson University, Clemson, SC 29634-09789}
\email{mbradle@clemson.edu}
\email{tlihsin@clemson.edu}

\altaffiltext{1}{Department of Physics and Astronomy, Kinard Laboratory of
Physics, Clemson University, Clemson, SC 29634-0978}

\begin{abstract}
We present a detailed study of the evolution of massive stars
of masses 15, 20, 25 and 30 $\msun$ assuming solar-like initial
chemical composition. The stellar sequences were evolved through
the advanced burning phases up to the end of core oxygen burning.
We present a careful analysis of the physical characteristics
of the stellar models. In particular, we investigate the effect of
the still unsettled reaction $^{12}$C($\alpha$,$\gamma$)$^{16}$O
on the advanced evolution by using recent compilations of this rate.
We find that this rate has a significant impact on the evolution
not only during the core helium burning phase, but also during
the late burning phases, especially the shell carbon-burning.
We have also considered the effect of different treatment of
convective instability based on the Ledoux criterion in regions
of varying molecular weight gradient during the hydrogen and helium
burning phases. We compare our results with other investigations
whenever available.
Finally, our present study constitutes the basis of analyzing
the nucleosynthesis processes in massive stars.
In particular we will present a detail analysis of the
{\it s}-process in a forthcoming paper.

\end{abstract}


\section{INTRODUCTION}

Study of the evolution of massive stars through the advanced burning phases
up to the collapse of the iron core is an active and complex field of
research in stellar astrophysics.
However, this does not mean that the
early hydrogen and helium burning phases are less important. Modeling stars is
an initial value problem, and as we shall see in this paper, these
early phases determine in a crucial way the initial conditions for the 
following burning phases, and the resulting stellar nucleosynthesis.
Therefore, despite several recent works dealing with the late evolution of 
massive stars 
(e.g. \citealp{1998ApJ...502..737C, 2000ApJS..129..625L, 
2002RvMP...74.1015W, 2001ApJ...558..903I}) 
an up-to-date, systematic study that also emphasizes the effect of
the early stellar phases on the later ones seems useful as well as a
discussion of uncertainties in key assumptions in the models.

The main emphasis of this paper, then, is an exploration of the physical 
conditions during the later burning phases of massive stars, how they
are set by earlier burning stages, and how they are affected by uncertainties
that remain in the input to the models.  A proper understanding of these
issues is important not only for clarifying the evolution of massive stars
but also for an accurate accounting of the stellar nucleosynthesis.

In this work, we present a systematic discussion of the 
evolution of massive stars through the H, He, C, Ne, and O burning
phases. We have made some effort to study the influence of uncertain
physical assumptions on the evolution.
In \S \ref{sec:stellarcode}, 
we give a brief description of the stellar evolution code
we have used in the present calculations. 
Section \ref{sec:characteristics} and its subsections
contain a detailed discussion of the characteristics of the 
burning phases mentioned above.
In \S \ref{sec:comparison}, 
we compare our results with those of other works.
Section \ref{sec:conclusions} contains our conclusions.

\section{STELLAR EVOLUTION CODE}
\label{sec:stellarcode}

The stellar evolution code used in this work is generally described
in \citet{2000ApJ...533..998T}.  As in our previous work,
the nuclear reaction network in the code contains 632 nuclear 
species, which allows us to follow most stellar nucleosynthesis processes,
including the {\it s}-process nucleosynthesis, in detail.  For the
present version of the code, we have updated the nuclear data.
We now use nuclear masses from compilation of 
\citet{1995NuPhA.595..409A} and
thermonuclear reaction rates from the 
compilation of the ``NACRE'' European collaboration 
\citep{1999NuPhA.656....3A},
and the ``Non-Smoker'' compilation of 
\citet{2000ADNDT..75....1R}.

In various stellar sequences described in 
\S \ref{sec:characteristics},
we have used
different reaction rates for the reaction 
$^{12}{\rm C}(\alpha,\gamma)^{16}{\rm O}$, which is still unsettled.
The latest compilation is due to \citet{2002ApJ...567..643K}. 
As stated in that work, the cross section of this reaction is
extremely small (10$^{-17}$ barn) at the energies relevant to core
helium burning; therefore, one has to rely on measurements 
at higher energies to obtain the S-factor at the relevant
stellar energies by extrapolation.

\citet{2002ApJ...567..643K}.
seem to have succeeded in reducing
the uncertainties in the evaluation of the cross section at stellar
temperatures. Fig. \ref{fig:C12agO16rate} 
shows clearly a different temperature
dependence of the rate obtained by \citet{2002ApJ...567..643K} compared to
other compilations.
Normalized to the rate of \citet{1988ADNDT..40..283C} (hereafter CF88),
we see that we are dealing with a different rate in the range of
helium burning (1.3-3.7$\times$10$^8$ K) and of carbon burning 
in the vicinity of T$_9$ = 1.
The implication of this new rate 
will be discussed in \S \ref{sec:characteristics} and \ref{sec:comparison}.

\section{CHARACTERISTICS OF THE BURNING PHASES} 
\label{sec:characteristics}

We present evolutionary sequences for stars of masses 
15, 20, 25, and 30 M$_{\sun}$ having solar-like initial 
chemical composition \citep{1989GeCoA..53..197A}.
All the evolutionary sequences have been evolved up to the end of
central oxygen burning.  Most models included mass loss
by stellar wind according to the semiempirical relation by
\citet{1988A&AS...72..259D}
and used the Schwarzschild criterion for convection throughout
the star.  In order to study the effects of mass loss,
the treatment of convective mixing in chemically inhomogeneous
layers, and the uncertainty in the 
$^{12}$C($\alpha$,$\gamma$)$^{16}$O reaction rate, we calculated
several evolutionary sequences with different input physics as 
summarized in Table 1. The only sequence calculated without mass
loss is for the 25 $\msun$ star (referred to as 25NM) in Table 1.
The case 25N is the same as 25NM, but mass loss is included here.
The case 25L is identical to 25N except that the Ledoux criterion
for convection was adopted in the region of the H-burning shell
(see \S \ref{sec:hydrogenhelium} for details). In the sequence 25K, the rate
for the reaction $^{12}$C($\alpha$,$\gamma$)$^{16}$O was used according
to \citet{2002ApJ...567..643K}, instead of the NACRE rate used in 25N.

Fig. \ref{fig:HRdiagram}a shows the HR diagram for our four evolutionary
sequences 15, 20, 25N, and 30  M$_{\sun}$. 
Physical characteristics of these tracks
are described in \S \ref{sec:hydrogenhelium}
with help of 
  Figs. \ref{fig:HRdiagram}b
and \ref{fig:Teff_vs_XHe4}.

Figs. \ref{fig:convective_zones_f15n0}-\ref{fig:convective_zones_f30n0}
 show the change of the internal structure of the stellar models at
a given time for all sequences listed in Table \ref{tab:stellarModels}.
In these figures we have labeled all convective zones according to their
physical origin, whether due to nuclear burning in a core or in shell region,
or in a star's envelope, or in an intermediate convection zone (referred to as 
ICZ) formed in the outer stellar layers above the receding convective core 
during core H-burning.
We will use these figures to describe the global characteristics of successive
burning phases as they occur in our evolutionary models.
Note that the ICZ in 
Figs.  \ref{fig:convective_zones_f15n0}-\ref{fig:convective_zones_f30n0}
contain the semiconvective and
convective regions that move in time due to the gradient in
hydrogen mass fraction and opacities. In our calculations we do not
include semiconvective mixing but apply the Ledoux criterion
in order to avoid introducing another parameter in our calculations
which is so far needed for describing semiconvective mixing.

\subsection{Hydrogen and Helium Burning Phases}
\label{sec:hydrogenhelium}
 
It is well known that significant changes of the 
evolutionary tracks occur during the hydrogen and helium
burning phases of massive stars;
therefore, we show in Fig. \ref{fig:HRdiagram}a
the evolutionary tracks in the HR-diagram for the stars under
consideration. 
In all models, we adopt the Schwarzchild criterion for convection
to determine the convective regions in a stellar model at a given 
time, except case 25L described below. The evolutionary 
tracks in Fig. \ref{fig:HRdiagram}a evolve nearly at 
constant luminosity after the main
sequence phase until they finally reach the red giant branch (RGB).
However, this evolution at constant luminosity and, more importantly,
the timing to reach the RGB depend crucially on mass loss and on how 
convective mixing is treated in the stellar layers of 
variable molecular weight gradient (or simply $\mu$-gradient) left
behind the shrinking hydrogen convective core during the main sequence evolution.

To understand these issues, we may compare different evolutionary sequences
as shown in Figs. 
 \ref{fig:HRdiagram}b
and \ref{fig:Teff_vs_XHe4}. 
  Fig. \ref{fig:HRdiagram}b
shows the evolutionary tracks when
mass loss is neglected as in our case 25NM, or when the convection treatment
is different as in case 25L. These cases are discussed below. 
Fig. \ref{fig:Teff_vs_XHe4} shows the effective temperature $T_{eff}$ as 
a function of the central helium mass fraction X($^{4}$He) and demonstrates
clearly how the evolution of the star to the RGB is influenced by mass loss.
In the case of no mass loss, the star remains in the blue region of the HR diagram
during most of its core He-burning. If mass loss is included, then the more 
massive the star, the stronger is the mass loss and the earlier is the transition
to the RGB. In other words, this transition occurs at higher central helium
mass fraction X($^{4}$He) (see Fig. \ref{fig:Teff_vs_XHe4})

To explain the behavior of the evolutionary tracks further, we compare
in detail the evolution of the 25 $\msun$ star with and without mass loss.
As Fig. \ref{fig:HRdiagram}b
shows, the sequence 25NM (without mass loss) has
higher luminosity during core hydrogen burning.
This is simply a consequence of the higher mass in this case.
During core helium burning, however, the explanation is more 
complicated since now two energy sources are available in
the star that contribute to the luminosity: 
a central source and shell source.
Why is the luminosity of the sequence with mass loss lower?.
We answer this question with the help of 
Figs. \ref{fig:convective_zones_f25n0}  and \ref{fig:convective_zones_25NM}
showing the change of internal structure in the course of evolution
for the sequences 25N and 25NM, respectively. 
In the early evolution through the hydrogen and
helium burning phases, these figures show a remarkably
different behavior of the convective zones
that forms above the hydrogen-burning shell.
In the sequence 25NM, this convective zone is more extended
in mass and lasts throughout the whole core helium burning
phase.  In contrast, in sequence 25N, calculated with mass loss, 
the convective zone is
narrower in mass and disappears before helium exhaustion in
the core (at X($^{4}$He) $\simeq$ 0.1).
The implication of this is that the H-burning shell in case
25NM is stronger because it is supplied by more fuel.

The stronger H-burning shell has several consequences
on the ensuing evolution:
\begin{enumerate}

\item It leads to a higher luminosity of the star during
core helium burning (see the tracks for the 25 M$_{\sun}$
star in Fig. \ref{fig:HRdiagram}b.

\item It delays the evolution to the RGB. The star remains
 as a blue supergiant during most of its core helium burning
 phase (case 25NM in Fig. \ref{fig:Teff_vs_XHe4}). 

\item It prolongs the duration of the core helium burning
 owing to the slightly lower central temperature in the
 case without mass loss (see Table \ref{tab:evolmod}).

\item It leads to a larger mass of the hydrogen exhausted
  core (or simply helium core) 
$M_{\alpha}$ = 7.62 M$_{\sun}$ in case 25NM compared to
$M_{\alpha}$ = 7.51 M$_{\sun}$ in case 25N
 (see Table \ref{tab:evolmod}).
\end{enumerate}

We point out that all these features have been recognized
by several authors 
(see \citet{1986ARA&A..24..329C}, for a review).
We refresh this issue, because it is relevant to our present discussion.

A more pronounced change of the structure of the convective zone formed
above the H-burning shell is found when the effect of the mean molecular
weight (or simply $\mu$-gradient) is taken into account to determine the
extent of this convective zone. It is well known (for a basic discussion
see \citealp{1990sse..book.....K}, \S 6) that the $\mu$-gradient acts to 
inhibit convection. In this case, convection is no longer a dynamical
instability, rather one deals with semiconvection, which is a consequence
of thermal instability. The criterion for convection that should be 
applied in this case is the ``Ledoux criterion'':
\begin{equation}
\nabla_{rad} \; > \; \nabla_{ad} \; + 
\frac{\varphi}{\delta} \; \nabla_{\mu}
\end{equation}
where the $\nabla$'s  are logarithmic gradients and the
quantities $\varphi$ and $\delta$ are derivatives obtained from
the equation of state (see \citealp{1990sse..book.....K}).
The second term $\nabla_{\mu}$ is always positive and
consequently inhibits convection. We will describe in the following
the results we have obtained for an evolutionary sequence of the 
25 $\msun$ star (labeled 25L in Table \ref{tab:stellarModels}),
where we have applied the Ledoux criterion only in the region 
of varying $\mu$-gradient, in which the H-burning shell is is effective.
Elsewhere in the stellar model, we have used the 
Schwarzschild criterion for convection.
We emphasize that our treatment maximizes the choking
effect of the $\mu$-gradient on convection.

As mentioned above, layers of varying $\mu$-gradient are
semiconvective. There is no general theory that describes the
efficiency of this process. When semiconvection is treated as a 
diffusion process (e.g., \citealp{1983A&A...126..207L, 1985A&A...145..179L,
1995MNRAS.275..983E, 2002RvMP...74.1015W} 
for a recent discussion), then a parameterization
of the diffusion coefficient for semiconvection is used.  
For our present investigation, we avoided using this
kind of parameterization by including the whole effect
of the $\mu$-gradient. We adopt the Ledoux criterion in regions of varying 
$\mu$-gradient in case 25L
and otherwise use the Schwarzschild criterion. A similar approach
has been done by \citet{1992ApJ...390..136S}.

In the following, we summarize the different features in the evolution
of the sequence 25L as compared to the case 25K during core helium burning.
The choking effect of the $\mu$-gradient on convection inhibits the formation
of a convection zone above the H-burning shell 
(compare Figs. \ref{fig:convective_zones_f25dx} and 
\ref{fig:convective_zones_f25n0}).
There are two key consequences of this. First the H-shell source is
weaker in 25L than in 25K, which explains the immediate evolution to the
 RGB as seen in Fig. \ref{fig:HRdiagram}b.
Secondly, owing to the weaker H-burning shell, the mass
of the convective core during core helium burning becomes 
larger in order to supply the luminosity of the star.
In 25K, the maximum convective helium core mass is 5.32 $\msun$
while in the case of 25L it is 6.16 $\msun$.
Therefore the H-shell is located farther out in mass in 25L.
The helium core mass is M$_{\alpha}$ = 7.54 $\msun$
for 25K, but 8.33 $\msun$ for 25L.

In addition to the above effects, there are
two other consequences of the weaker H-shell.
First, the
larger mass of the convective core leads to 
a shorter
duration of core helium burning due to the higher
central temperature: 6.88$\times$10$^5$ yrs for 25L
compared to 7.49$\times$10$^5$ yrs for 25K.
Second, in the case of a 
weaker H-shell, 
the He-shell burning occurs in a region
M$_r$ = 6.184 - 8.108 $\msun$, while in case 25K
this region comprises M$_r$ = 5.393 - 7.290 $\msun$.
In other words, when the Ledoux criterion is used in
the H-shell region, the He-burning shell migrates
farther out by about 0.8 $\msun$ compared to the
case with Schwarzschild criterion everywhere in the
stellar model.

It is interesting to note that our sequence 25L has many similarities
with the sequence of the 25 $\msun$ star calculated by
\citet{2002RvMP...74.1015W} (hereafter WHW).
We obtain close agreement with these authors concerning
the depth of the outer convective envelope and the
He-burning shell.
We achieve good agreement concerning the higher total
mass lost from this star, because in both calculations
the stellar sequences evolve directly to the RGB,
after the main sequence phase, where mass loss increases
significantly.  Finally, we also find agreement for the
mass of the helium core 
(see Table \ref{tab:compmod25} for more details).

Summing up this discussion, then, the treatment of
convective mixing in the region of varying 
$\mu$-gradient where the H-burning shell is active
affects many physical quantities relevant to
the evolution past core helium burning.

In connection with the early evolution phases, we
may also comment on the effects of mass loss 
and envelope convection on the surface abundances,
which are modified when the stellar sequences reach the RGB and
develop convective envelopes, such that the processed 
material is mixed to the surface (the well-known "first dredge up").
Table \ref{tab:SurfAbundRatio} shows the surface abundances as they
result after the first dredge up for the stars under consideration. 
The ratio of $^{12}$C/$^{13}$C decreases with 
increasing stellar mass (15, 20, 25N, and 30) owing
to the larger production of $^{13}$C by the
CNO cycle in more massive stars.
When mass loss is neglected as in case 25NM, 
this ratio becomes larger.

The ratio $^{14}$N/$^{15}$N increases strongly with
the stellar mass mainly due to the destruction of
$^{15}$N by the (p,$\alpha$) reaction in the CNO cycle.
At constant mass, $^{15}$N is strongly reduced because
convective mixing extends over a larger mass of
the envelope. Eventually, the ratio $^{14}$N/$^{15}$N
may be used as indicator of mass loss.

The ratio $^{16}$O/$^{18}$O is an increasing function
of the stellar mass. Its behavior is similar to 
$^{14}$N/$^{15}$N, although its variation is not so strong.

In contrast to the behavior discussed above, the
ratio $^{16}$O/$^{17}$O does not have a monotonic
dependence on the stellar mass.
This is because the production and destruction of
$^{17}$O is rather sensitive to temperature 
variations influencing the reaction
$^{17}$O(p,$\alpha$)$^{14}$N in particular.
The effect of mass loss leads to a slight decrease 
of this ratio.
The enrichment of sodium compared to solar
([Na/Fe] in Table \ref{tab:SurfAbundRatio}) 
is an increasing function of the stellar mass owing to 
the more efficient Ne-Mg cycle in more massive stars.

It may be also interesting to compare the surface isotopic ratios
of the the sequence 25L with those of 25N. As shown in Table 2,
a significant difference is found for the ratios  $^{14}$N/$^{15}$N,
$^{16}$O/$^{18}$O and [Na/Fe]. The isotopic ratios
of the sequence 25N represent the products of shell
hydrogen-burning which was operating at higher temperatures (by 
a factor of about two at the bottom of the shell) than in case 25L.
This is clearly seen by the higher enrichment of Na, and the high 
destruction of $^{15}$N and  $^{18}$O. These products are mixed to the
surface from deeper stellar layers in case of 25N, where the convective
envelope penetrates down to 7.6 $\msun$ compared to 8.7 $\msun$ in case
25L. We observe that the surface isotopic ratio in massive stars may
depend on how convection is treated in the layer where shell hydrogen burning 
proceeds.

Table \ref{tab:evolmod} presents a few more details about 
the stellar sequences listed
in Table \ref{tab:stellarModels}. 
As may be  seen, the final masses of the 
helium cores, $M_{\alpha}$ of the stars under consideration are
3.85 $\msun$ for the 15 $\msun$,
5.70 $\msun$ for the 20 $\msun$,
7.53 $\msun$ for the 25 $\msun$ with mass loss,
and 
9.54 $\msun$ for the 30 $\msun$.
As Table \ref{tab:evolmod}, 
indicates, M$_{\alpha}$ ceases to change
at the end of core carbon burning.
Our values above are in close agreement with the results
by \citet{1988PhR...163...13N} and \citet{1996snai.book.....A}.
However, we think that our results are more accurate
because those authors followed
the evolution of helium cores only, that is without 
including the H-burning shell in their calculations.
We make another comparison with other authors in 
Table \ref{tab:compmod25} (see \S \ref{sec:comparison}).

We conclude this section by discussing a very important 
physical quantity determined at the end of 
of core He-burning, namely, the central mass fraction of carbon , X($^{12}$C).
Table \ref{tab:COMassfrac} shows that using the \citet{2002ApJ...567..643K}
reaction rate for $^{12}$C($\alpha$,$\gamma$)$^{16}$O leaves a larger 
X($^{12}$C) than using the NACRE rate.  As Fig. \ref{fig:C12agO16rate} shows,
the \citet{2002ApJ...567..643K} rate is lower than 
the NACRE rate for $T_9 < 0.4$,
which explains why the former rate allows for less destruction of 
$^{12}$C during the late stages of core helium burning. 
Interestingly, among the 25 $\msun$ models,
the lowest value of X($^{12}$C)--0.193--is
obtained for the 25NM.  This is due to the higher central
temperature encountered in this case which leads to more efficient
destruction of $^{12}$C.  
The central X($^{12}$C) strongly affects the subsequent
evolution of the star \citep{1993ApJ...411..823W, 2001ApJ...558..903I}.

\subsection{Carbon Burning Phase}
\label{sec:carbonphase}

In this section, we present some details about core and shell
carbon burning.  
The evolution of the centers of the stars of masses 15 to 30 $\msun$
in a ${\rm T}_C-\rho_{\rm C}$ diagram is shown in Fig. \ref{fig:Tc_vs_rhoC}
where the ignition temperatures of the calculated burning phases 
are indicated by the dashed lines (see entry 5 in Table \ref{tab:evolmod}).
Core carbon burning occurs in the temperature range (0.32 - 1.2)$\times$10$^9$ K 
where the lowest value belongs to the 15 $\msun$ star while the highest value
belongs to the 30 $\msun$ star. 

Core carbon burning in the 15 $\msun$ star occurs
in a convective core whose mass reaches a maximum value
of 0.55 $\msun$. The duration of core carbon burning is 6760 yrs.
When the carbon is exhausted in the core, shell-carbon burning
mediates the evolution toward neon burning; in contrast to the 
helium-burning shell that has its location in mass radius fixed at all
times (see Figs. 
\ref{fig:convective_zones_f15n0} - \ref{fig:convective_zones_f30n0}),
carbon burning in a shell exhibits complicated behavior.
In the 15 $\msun$ model, two episodes of convective  shell C-burning occurs
before Ne-ignition where the shell advances in mass 
(see Fig. \ref{fig:convective_zones_f15n0}).

Core carbon burning in the 20 $\msun$ star occurs in a
convective core whose mass reaches a maximum of 0.49 $\msun$.
The duration of this phase is 3300 yrs.
The shell C-burning proceeds here also in two convective episodes.  We note that
during the second episode, a broader convective zone is formed than
in the 15 $\msun$ model (see Fig. \ref{fig:convective_zones_f20n0}).

In the 25 $\msun$ model (case 25N), core carbon burning occurs in a
convective core whose mass reaches a maximum  of 0.35 $\msun$. 
In this case, there are three subsequent carbon shells. The last of these
has a range of
(2.26 - 4.54)$\msun$ and persists to the end of the calculation.
How the shell C-burning proceeds depend crucially on the carbon profile
created in the star after core C-burning. 
  Fig. \ref{fig:s25k0_massfrac_vs_Mr_mod05050_EndofNeburn}a
  and  \ref{fig:s25k0_massfrac_vs_Mr_mod05050_EndofNeburn}b
show clear difference
in such profile. 

It is interesting to compare the effect of different rates
for the reaction $^{12}$C($\alpha$,$\gamma$)$^{16}$O on carbon burning in
the 25 $\msun$ models.
Comparison of Figs. \ref{fig:convective_zones_f25n0} and
\ref{fig:convective_zones_f25l0}
shows the structural evolution of the stellar sequences
25N and 25K as obtained from the Schwarzschild criterion for convection
and the different rates of $^{12}$C($\alpha$,$\gamma$)$^{16}$O as in Table 1.
As we saw previously in Table \ref{tab:COMassfrac}, X($^{12}$C)=0.236
is left at the end of core helium burning when the rate of  
$^{12}$C($\alpha$,$\gamma$)$^{16}$O according to NACRE is used, but
X($^{12}$C) = 0.280 according to \citet{2002ApJ...567..643K}.
The larger X($^{12}$C) at the end of core helium burning
allows 25K to develop a convective core with a maximum mass of 0.47 $\msun$
in contrast to 0.36 $\msun$ for 25N.

It seems that the higher the value of $^{12}$C, the larger is the mass 
of the convective core during core C-burning. This is due to the
well-known interplay between the energy generation rate and neutrino loss 
rate 
\citep{1972ApJ...176..699A, 1986ARA&A..24..205W, 
1998ApJ...502..737C, 2000ApJS..129..625L}.
A convective core forms only when the effective
energy generation rate (nuclear minus neutrino) 
is positive.
The difference in the mass of the convective core has
a consequence on the duration of core C-burning: it lasts 3500 yrs 
in case 25K, but 1600 yrs in case 25N. 

It is also interesting to compare 25N to our
sequence without mass loss designated as 25NM in
Table \ref{tab:stellarModels} and Fig. \ref{fig:convective_zones_25NM}.
At the end of core He-burning, X($^{12}$C) = 0.193. 
The low value of X($^{12}$C) leads to
a maximum mass of 0.23 for the convective core
during core C-burning in 25NM, but a duration for
this phase of 1860 yrs, longer than in 25N.
While the mass of the convective core during core C-burning
is correlated to X($^{12}$C) at the beginning of this
burning phase, the lifetime of this phase has no simple
relation to X($^{12}$C), because it depends also sensitively
on how the center of the stars evolve in 
a ${\rm T}_C$-$\rho_C$ plane.
Fig. \ref{fig:Tc_vs_rhoC_25msun}a 
shows this kind of evolution is rather complicated,
which affects the nuclear energy generation rates and the 
neutrino loss rates leading to the variation of the
lifetime found above.

From Fig. \ref{fig:convective_zones_f30n0} and Table \ref{tab:evolmod},
we see that the 30 $\msun$
model has a maximum carbon-convective core mass of 0.39 $\msun$.  This mass
is somewhat larger than that in 25N, even though the mass fraction
of $^{12}$C is larger in 25N than the 30 $\msun$ model, an
exception to the general rule that more carbon gives a larger
convective carbon core.  On the other hand, we note that at the time
of core carbon burning, the 30 $\msun$ star has in fact
a lower mass than that of 25N.  This and the complicated interplay
between energy generation and neutrino energy loss obscure a direct 
relation between the convective core mass and X($^{12}$C).

We now focus in more detail on shell C-burning, and consider in particular
the 25 $\msun$ sequences.
As seen in Figs. \ref{fig:convective_zones_f25n0} 
and \ref{fig:convective_zones_f25l0},
the shell C-burning starts in a first phase near the edge
of the former convective core with a convective zone formed
above the shell. A second convective phase of this shell burning
comprises more mass in the sequence 25K than in 25N.
How C-shell burning proceeds depends crucially on the
carbon profile left in the star after core C-burning.

Fig. \ref{fig:s25k0_massfrac_vs_Mr_mod05050_EndofNeburn}a
and  \ref{fig:s25k0_massfrac_vs_Mr_mod05050_EndofNeburn}b
help to illustrate this point.
They show the composition profile of major elements
at the end of core Ne burning. The profile of carbon
at this stage is clearly different. 
In the sequence 25K, the second convective phase
of C-shell burning comprises more mass compared to
that of the sequence 25N.
Consequently, C-shell burning in 25K is stronger and 
leads to a significantly reduced carbon abundance in the
mass range 1.5 - 4.0 $\msun$ 
(see Fig. \ref{fig:s25k0_massfrac_vs_Mr_mod05050_EndofNeburn}a
as compared to 
Fig. \ref{fig:s25k0_massfrac_vs_Mr_mod05050_EndofNeburn}b.
Clearly, the rate of $^{12}$C($\alpha$,$\gamma$)$^{16}$O
has a significant effect on the properties of the shell C-burning.


To complete the discussion, we point out that a 
convective core of 0.55 $\msun$ is formed 
 (see Fig. \ref{fig:s25k0_massfrac_vs_Mr_mod05050_EndofNeburn}b)
in the 
15 $\msun$ sequence during core C-burning and that shell
C-burning proceeds in two convective episodes prior to
Ne ignition.
By contrast, in the 30 $\msun$ sequence, the mass of the 
convective core is 0.39 $\msun$ 
(see Fig. \ref{fig:convective_zones_f30n0}) and 
C-shell burning constitutes only one single convective episode.

When the core C-burning ends and the stars contract toward
core Ne-ignition, a third convective C-shell episode starts
in all our cases except in the 30 $\msun$ sequence.
This third convective C-shell lasts all the time as far as
we have carried out calculations beyond core O-burning.
Indeed this last phase of C-shell burning seems to survive
till core collapse as it is seen in the calculations carried 
to the collapse phase \citep{1998ApJ...502..737C,
2000ApJS..129..625L, 2002RvMP...74.1015W}.

\subsection{Neon Burning Phase}
\label{sec:neonphase}

For the stars under consideration, core Ne-burning
occurs in the temperature range (1.2 - 1.9)$\times$10$^9$K
(see entry 7 in Table \ref{tab:evolmod}). 
A characteristic of this burning phase is that a convective
core is always formed in the mass range of 0.6 - 0.75 $\msun$,
but it exists only during a short time interval during which 
the effective energy generation rate (nuclear - neutrino) 
is positive; thus, core Ne-burning is partially radiative and
of short duration (3.13 yrs for the 15 $\msun$, 
1.21 yrs for the 20 $\msun$, 0.33 yrs for the 30 $\msun$).

The effect of the rate for the reaction 
$^{12}$C($\alpha$,$\gamma$)$^{16}$O
is also encountered during this phase.
Comparing the sequences 25K and 25N, we find that the
convective core attains a maximum mass of 0.61 $\msun$ for 25N,
but 0.75 $\msun$ for 25K.
The lifetime of core Ne-burning is rather different:
0.294 yrs for 25N, but 3.28 yrs for 25K.

The shorter lifetime of the sequence 25N is not only
a consequence of the relatively smaller mass of its
convective core, but it can also be traced back to the
effect of 
$^{12}$C($\alpha$,$\gamma$)$^{16}$O.  Sequence 25K achieves
a larger central mass fraction of Ne at the end of core
carbon burning (0.384, see Table \ref{tab:COMassfrac}).
As Figs. \ref{fig:Tc_vs_rhoC_25msun}a and
 \ref{fig:Tc_vs_rhoC_25msun}b
show, this larger Ne mass fraction allows
the sequence 25K to evolve at relatively lower central
temperatures but higher densities during core Ne-burning than the sequence 25N.
It also results in a pronounced expansion phase during core
Ne burning, as seen in Fig. \ref{fig:Tc_vs_rhoC_25msun}a, 
The longer lifetime of core Ne-burning in 25K than in 25N
is thus due to fact that 25K has more Ne to burn and does so
at relatively lower temperatures.

In the 25 $\msun$ models, the third
convective phase of C-shell burning occurs during 
core Ne-burning.  A significant difference exists
in this phase between 25K and 25N.  In 25K,
the convective zone above the C-shell is quite narrow
in mass.  This is due to the fact that
the mass fraction of $^{12}$C is relatively high throughout
the helium exhausted core; therefore, when carbon
burning occurs during the second convective phase
of the C-shell, the burning is vigorous, which gives
an extended second carbon convective shell and largely
depletes the $^{12}$C.  This makes less carbon available
for the third convective carbon shell, which, consequently
weaker and narrower.
It is interesting to see that the effect of the rate for
$^{12}$C($\alpha$,$\gamma$)$^{16}$O
propagates so effectively into the advanced burning phases of
massive stars.

\subsection{Oxygen Burning Phase}

For our stars, core oxygen burning (or O-burning) starts
in the temperature range of (1.5 - 2.6)$\times$10$^9$ K
(see entry 9 in Table \ref{tab:evolmod}).
One common characteristic of our stars when they evolve 
to this stage is that they develop convective cores of
mass in the range of (0.77 - 1.20) $\msun$ as can be
seen in Table \ref{tab:evolmod}.
These relatively high masses result from the fact that
all our sequences have high central mass fractions 
of $^{16}$O after completing core Ne burning.
Consequently the effective energy generation rate
(nuclear - neutrino) is strongly positive throughout this
burning phase and proceeds in a convective core
(see Figs. \ref{fig:convective_zones_f15n0} to \ref{fig:convective_zones_f30n0}).

We observe from Table \ref{tab:evolmod} (entry 10) that the lifetime
of core oxygen burning generally decreases monotonically
with stellar mass.  Higher mass stars typically burn
at higher central temperatures for which the
$^{16}{\rm O}$ + $^{16}{\rm O}$ reaction is faster.
This also explains the difference in lifetimes of
core oxygen burning in 25K and 25N.  We see
from  Table \ref{tab:evolmod}
that the former model has a longer core oxygen
burning lifetime.  This is because 
it evolves at lower central temperature during the early phase
of core oxygen burning (see Fig. \ref{fig:Tc_vs_rhoC_25msun}a ).
Again, we trace back the effect of different rates for
$^{12}$C($\alpha$,$\gamma$)$^{16}$O
in these cases.

After core oxygen burning, our models typically show a brief shell neon
burning phase followed by a shell oxygen phase.  Both of these shells
are relatively narrow in extent. Our calculations were not carried 
beyond shell oxygen-burning. 

Finally, we show in Fig. \ref{fig:dns_vs_Mr_at_end_Oburn}
the density and temperature profiles
inside our stars as far as we have evolved them,
mainly beyond core oxygen burning.
The jumps in density seen in Fig. \ref{fig:dns_vs_Mr_at_end_Oburn}a
indicate clearly the different cores of the star,
where the main fuel of a burning phase
is exhausted.
Our Fig. \ref{fig:dns_vs_Mr_at_end_Oburn}a is rather similar to 
Fig. 12 obtained by
\citet{2000ApJS..129..625L}.

\section{COMPARISON WITH OTHER INVESTIGATIONS} 
\label{sec:comparison}

It is worthwhile to compare our results with those
obtained by other groups with the aim to find out
where a close agreement can be achieved and under which
physical conditions.
This will help in understanding the complex nature of 
the advanced burning phases of massive stars and the
role of key physical assumptions in the stellar models.
Such a comparison, which we present in
Tables \ref{tab:compmod15} and \ref{tab:compmod25},
is laborious owing to the many parameters involved.
Nevertheless, it allows us to consider in detail
the three important parameters
we have investigated: mass loss, treatment
of convection in layers of varying $\mu$-gradient, and 
the rate of $^{12}$C($\alpha$,$\gamma$)$^{16}$O.

The mass fraction of $^{12}$C left over at the end of core helium
burning depends on the  $^{12}$C($\alpha$,$\gamma$)$^{16}$O rate 
and the treatment of the convection.
When comparing our results to those obtained by 
\citet{2000ApJS..129..625L} (hereafter LSC) in 
Table 5 \& 6, 
we note that they have evolved their stars at
constant mass, used the rate of
\citet{1985ADNDT..32..197C} 
(hereafter CFHZ85) for
$^{12}$C($\alpha$,$\gamma$)$^{16}$O,
and treated convection on the basis of the Schwarzschild
criterion.
For the 15 $\msun$ star, Table \ref{tab:compmod15} shows
that our results are generally in reasonable
agreement with theirs 
where both models apply the same criterion for convection.
There is a difference in the central mass fraction
of carbon X($^{12}$C) in Table 5 \& 6) left over at the end of core
helium burning: 0.285 in our case compared to 0.195 according to LSC.
The reason for our higher values is due to
the lower rate of $^{12}$C($\alpha$,$\gamma$)$^{16}$O
we have used (NACRE's compilation,
see Fig. \ref{fig:C12agO16rate}). 
The higher value of X($^{12}$C) in our case is also responsible
for the difference in the lifetime of core C-burning and to the higher mass
of the convective core we find.
Also the difference in the location and extension 
of the C-burning shell in our case is also attributable
to the higher value of X($^{12}$C) which determines
the carbon profile in the region of the C-shell.
A close agreement is found for the Ne-burning phase.
For the core O-burning we find a slightly longer lifetime for 
our 15 $\msun$ sequence.

Our comparison with the results of
WHW in Table \ref{tab:compmod15} \& \ref{tab:compmod25}
reveals that we can present a similar comparison to
what we have done for LSC.
We note, however, that WHW have used the rate of 
$^{12}$C($\alpha$,$\gamma$)$^{16}$O
obtained by \citet{1996ApJ...468L.127B}
but multiplied by a factor of 1.2 in their calculations.
In Fig. \ref{fig:C12agO16rate}, we see that this is 
comparable to the rate due to \citet{2002ApJ...567..643K},
which is lower than the NACRE rate
we have used in the 15 $\msun$ star for T$_9 <$ 0.4.
WHW also use a relatively large semiconvective diffusion
coefficient implying overshooting, semiconvection, and
rotation-induced mixing.
Therefore, we think that the relatively low value
X($^{12}$C) = 0.219 obtained by WHW for the 15 $\msun$
at the end of core He-burning 
is due to the combined effect of both the 
$^{12}$C($\alpha$,$\gamma$)$^{16}$O 
and the treatment of convection.
We may also relate the differences for the location and extent
of the C-burning shell compared with WHW to the low value of 
 X($^{12}$C) in their model. This also explains the differences
during core Ne-burning. Close agreement is found for the core O-burning.

In the case of the 25 $\msun$ sequence, our comparison
is presented in Table 6 including four of our sequences (25K, 25N, 25NM, 25L).
In this case, mass loss by stellar wind is much
higher than in the case of the 15 $\msun$ sequence,
as may be seen in the second column of 
Table \ref{tab:evolmod}.
Our comparison with the results by LSC will be done
with 25NM, which, like the LSC model, was evolved at constant mass.
On the other hand, our sequence 25N evolved
with the same input physics but included mass loss
should reveal the effect of that parameter on the evolution.

We find close agreement between our sequence 25NM and
that by LSC during the H-burning and He-burning phases.
A slight difference is found for the mass of the
helium core and the depth of the convective envelope.
The quantities listed in Table \ref{tab:compmod25}
for core He-burning are in good agreement with those
by LSC, including the X($^{12}$C).
As expected, the differences are more significant
when we compare with our sequence with 
mass loss (25N).
Significantly, the value of X($^{12}$C) = 0.193 in
25NM is lower than 0.236 in 25N.
The reason for this is that, toward the end of core He-burning,
or, more precisely, when X($^{4}$He) $\leq$ 0.10,
the star has evolved
to the red giant branch where mass loss increases
significantly, as seen in Fig. \ref{fig:Mass_vs_XHe4_25Msun_Heburn}).
As a consequence, the star evolves at lower central temperature
because it is now a lower mass star.
The net result is the high value of X($^{12}$C) mentioned above.

\citet{2001ApJ...558..903I} have studied the influence of
different rates of $^{12}$C($\alpha$,$\gamma$)$^{16}$O
reaction in massive stars.
They used the rate of CF88 and that of CFHZ85, which is higher
by a factor of $\simeq$2.7. The CF88 rate led to a central mass fraction of 
carbon X($^{12}$C) = 0.424 at the end of core
He-burning, while the second rate led to
X($^{12}$C) = 0.20.
As we have seen, however, when we compare our results
for our sequence 25NM with those of LSC,
we find close agreement in X($^{12}$C)  
(0.19 in our case to 0.18 in their case).
The slight difference is mainly due to the fact
that the rate of
CFHZ85 for $^{12}$C($\alpha$,$\gamma$)$^{16}$O
used by LSC is about a factor of 1.4$\times$ larger
than the NACRE rate we have used in 25NM.

We obtain a longer lifetime for the
carbon core phase in 25NM than in LSC's 25 $\msun$ model
because in 25NM it occurs partly in a convective core of 0.23 $\msun$, 
which is not present in the case of LSC.
During core carbon burning phase, it is known that the
neutrino energy losses influences significantly the evolution of
the central temperature and density of massive stars. 
Perhaps the difference in the neutrino loss prescription
is partly the
cause of the difference in the carbon core evolution.
Our neutrino energy losses due to pair, photo, plasma,
bremsstrahlung, and recombination processes are taken from
the analytical approximation of 
\citet{1996ApJS..102..411I}.
LSC adopted the photo, pair, and plasma neutrino energy losses
of \citet{1985ApJ...296..197M} and \citet{1986ApJ...304..580M},
while their bremsstrahlung neutrino energy loss is taken
following 
\citet{1976ApJ...210..481D} and 
\citet{1982ApJ...255..624R}
and their recombination neutrino energy loss follows the
prescription of 
\citet{1967ApJ...150..979B}.
Another contributing factor is
the larger value of X($^{12}$C) at the end of core helium burning
in 25NM than of LSC's 25 $\msun$ model which is due to
the smaller rate for $^{12}{\rm C}(\alpha,\gamma)^{16}$O used in 25NM.
The presence of a convective core in our case influences strongly
the characteristics of the ensuing shell carbon-burning.
In the case of 25NM, our first convective carbon-burning shell is
located deeper in the star as compared with 
LSC (see Table \ref{tab:compmod25}).
The reason is that the presence of a convective core in our case
leads to to the formation of a steep carbon gradient near to
its edge, such that the high carbon abundances drives convection.
Note that we have close agreement with LSC concerning the locations
of the convective carbon-burning shell in the case of the
15 $\msun$ star where a convective core is found in both calculations.

Among our 25 $\msun$ models, it is most useful to compare 25K to the 
25 $\msun$ model of WHW.
This is because WHW include mass loss and because
we have used the rate for 
$^{12}$C($\alpha$,$\gamma$)$^{16}$O
for this sequence that is closest to the rate used by WHW
(based on Buchmann's compilation as mentioned above) 
in the temperature range of core helium burning 
(see Fig.  \ref{fig:C12agO16rate}). 

The remarkable differences are:
the lifetime of core H-burning and the mass of the
He-core are larger in WHW's sequence.
The first indicates a certain degree of overshooting
in the prescription of convection and 
the second is due to semiconvection being used in
the region of the H-burning shell.
In \S \ref{sec:hydrogenhelium}, we have described this issue
and found close agreement between the calculations by WHW
and our results for the sequence 25L.
We recall that in this sequence the inhibiting effect of
the $\mu$-gradient on convection was taken into account
in the region of H-burning shell.

We emphasize that the convective He-burning shell in our
sequence 25L (see Fig. \ref{fig:convective_zones_f25dx})
has a similar location to that in the
calculations of WHW for their 25 $\msun$ star.
Compared to our sequence 25K, this zone is shifted 
by 0.8 $\msun$ outward.
This is a consequence of the larger convective core during
helium burning.
We note that we still have a higher value of X($^{12}$C)
at the end of core helium burning
than obtained by WHW for similar reasons as explained
in the case of the 15 $\msun$ above.

During the core C-burning, our calculations show that
a convective core is formed with maximum mass that increases
with increasing value of X($^{12}$C).
There is no convective core formed in the calculations
by WHW, and this explains the shorter lifetime of 
C-burning they have obtained as compared with ours.
Whether a convective core is formed or not during core
C-burning is crucially dependent on the balance between
the nuclear energy generation rate and the neutrino
loss rate.  It is clear that the former is very sensitive
to the value of X($^{12}$C) retained at the end of core
He-burning.
The manner in which core C-burning proceeds will of course
determine how carbon-shell burning proceeds later.
This explains the different locations and extension
of the C-burning shell in our calculations.
Indeed, if we go back to the 15 $\msun$ sequence in
Table \ref{tab:compmod15} and
compare our results for the C-shell burning, we find
reasonable agreement with others, because all the
calculations presented in 
Table \ref{tab:compmod15} have a convective core during core C-burning.
The fact that the more massive star such as 25 $\msun$ which has
higher values of temperature and density, and lower degeneracy
than the 15 $\msun$ star make the effect of neutrino energy losses
more pronounce. Therefore, the different prescription of the
neutrino energy losses makes the carbon core evolution differences
of the 25 $\msun$ models more evident.

The existence or absence of a convective core during
core C-burning will also influence the properties
of the Ne-burning phase.
It is interesting that the oxygen-burning phase
is more robust, such that a reasonable agreement is
achieved among different calculations as seen
in Table \ref{tab:compmod15} and \ref{tab:compmod25}.
This is likely due to the dominance of the
nuclear energy production over the neutrino
losses during the core O-burning phase.

Summing up this discussion, it is clear that the
evolution of massive stars through the advanced
burning phases (core C-burning and beyond) is quite
sensitive to the earlier evolution.  This, in turn,
means that the nature of the burning in the models
in the advanced phases depends strongly on the
the effects of mass loss, on the treatment of
convection in inhomogeneous stellar layers, and on
the central carbon mass fraction retained at the
end of core He-burning.

\section{CONCLUSIONS}
\label{sec:conclusions}

In this work, we have evolved models of stars of mass
15 - 30 $\msun$ through most of their burning phases.
We have analyzed the effects of three important physical
ingredients on the structures of the stellar models, viz.,
mass loss by stellar wind, the recently suggested rates for the 
$^{12}$C($\alpha$,$\gamma$)$^{16}$O
reaction, and
the treatment of convection in inhomogeneous stellar layers
on the basis of the Ledoux criterion in contrast to the
Schwarzschild criterion.
We summarize the main results as follows:

\begin{itemize}
\item Mass loss has a strong effect on stellar evolution through
advanced phases.  The larger the star is initially, the
more mass loss will decrease the stellar mass.
Then the star, having lower mass, will evolve at lower central
temperatures and densities, such that less carbon would be
destroyed by 
$^{12}$C($\alpha$,$\gamma$)$^{16}$O
during core He burning.
On the other hand, Figs. \ref{fig:Teff_vs_XHe4} and
\ref{fig:Mass_vs_XHe4_25Msun_Heburn}
show that mass loss increases strongly when the star
reaches the red giant branch (RGB).



\item Our discussion of the sequence 25L in \S \ref{sec:hydrogenhelium}
has  emphasized that mass loss depends sensitively on the behavior
of the hydrogen-burning shell during core He-burning.
When convection is inhibited in the hydrogen shell region by
including the chocking effecdt of the $\mu$-gradient,
then the star evolves directly to the RGB, such that it looses
more mass.
The effect of this convection treatment was also that the 
helium burning shell and the hydrogen-burning shell migrate
outward in mass and burn at relatively lower temperatures.

\item Our evolutionary sequences 25K, 25N, and 25NM indicate
that the central mass fraction of carbon, X($^{12}$C) left
after core helium burning determines the physical characteristics
of the ensuing core carbon burning and shell carbon burning.
The larger X($^{12}$C) is, the larger is the mass of the 
convective core during core carbon burning.
The formation of such a core in our calculations leads to
a steep gradient of carbon at the edge of the core such
that shell carbon-burning occurs first in a convective zone
located deeper in the star compared to other 
calculation in which no convective core is
formed (see Table \ref{tab:compmod25}).

\item Core Ne-burning is also influenced by the 
characteristics of the preceding burning phases as described
in \S \ref{sec:neonphase}.

\item Core oxygen burning is found to be less sensitive to the
preceding burning phases, because the oxygen central mass fraction
at the end of core neon-burning is always high.
This leads to the dominance of the nuclear energy production
over neutrino energy losses, such that a general agreement
is found among different calculations 
(see Table \ref{tab:compmod15} and \ref{tab:compmod25})
\end{itemize}

From this work, we see the complex sensitivity of the
structure and evolution of massive stars to mass loss,
convection, and the $^{12}{\rm C}(\alpha,\gamma)^{16}$O
rate.  An exact treatment of these three key ingredients
to the models is not yet available; therefore, one must remain aware of their
effects on stellar models.  This becomes especially important
when one considers the resulting nucleosynthesis, as
we will do in subsequent work.

\acknowledgments

We are grateful to Donald Clayton 
for helpful comments and suggestions.
This work has been supported by NSF grant AST-9819877
and by grants from NASA's Cosmochemistry Program
and from the DOE's Scientific Discovery 
through Advanced Computing Program.
M. F. El Eid is grateful to the American University of Beirut (AUB)
for a paid research leave and to Clemson University for support and hospitality.

\bibliographystyle{apj}

\clearpage


\begin{deluxetable}{cccc}
\footnotesize
\tablecolumns{4}
\tablewidth{0pc}
\tablecaption{
List of Stellar Sequences Studied}
\tablehead{
\colhead{Star Mass (M$_{\sun}$)} & Massloss &
                 \colhead{$^{12}$C($\alpha$,$\gamma$)$^{16}$O} &
                 \colhead{$^{22}$Ne($\alpha$,n)$^{25}$Mg} 
}
\startdata
15       & yes & NACRE$^a$ & NACRE \\
20       & yes & NACRE & NACRE \\
25K      & yes & Kunz$^b$  & NACRE \\
25L$^c$  & yes & Kunz  & NACRE \\
25N      & yes & NACRE & NACRE \\
25NM     & no  & NACRE & NACRE \\
30       & yes & NACRE & NACRE 
\enddata
\tablenotetext{a}{Rates according to Angulo et al. (1999).}
\tablenotetext{b}{Rates according to Kunz et al. (2002).}
\tablenotetext{c}{This sequence is obtained by adopting the Ledoux
criterion for convection in the regions of the H-burning shell
(see Chap. 3.1)}
\label{tab:stellarModels}
\end{deluxetable}
\clearpage

\begin{deluxetable}{lccccccc} 
\footnotesize 
\tablewidth{0pt} 
\tablecaption{Surface Abundance Ratios (by numbers)}
\tablehead{ 
\colhead{} & \colhead{15}   & \colhead{20} &
\colhead{25N}  & \colhead{25NM}  & \colhead{25L} & \colhead{30} & \colhead{Initial Value}
}
\startdata
$^{12}$C/$^{13}$C   &  18.2   &  17.8   &  15.6   &  28.9    & 17.49   &  7.94   & 90.0 \\
                    &  18.4   &  16.4   &  8.63   & \nodata  & \nodata & \nodata & \\
                    &  93     &  91     & \nodata &  90      & \nodata & \nodata & \\
$^{14}$N/$^{15}$N   &  2489   &  3416   &  7140   &  3334    &  2469   & 20681   & 271  \\
                    & \nodata & \nodata & \nodata & \nodata  & \nodata & \nodata &      \\
                    &  2132   &  2702   & \nodata &  2977    & \nodata & \nodata &      \\
$^{16}$O/$^{17}$O   &  930    &  1054   &  1022   &  1366    &  1309   &   699   & 2622 \\
                    &  1231   &  1433   &  1159   & \nodata  & \nodata & \nodata &  \\
                    &  881    &  919    & \nodata &  1052    & \nodata & \nodata &  \\
$^{16}$O/$^{18}$O   &  686    &  705    &  846    &   691    &  614    &  1976   & 498  \\
                    &  579    &  650    &  1227   & \nodata  & \nodata & \nodata &  \\
                    &  565    &  574    & \nodata &  572     & \nodata & \nodata &  \\
$[$Na/Fe$]$         &  0.271  & 0.361   &  0.494  &  0.451   &  0.302  &  0.617  &  0   \\
M$_{bottom}$/M$_{\sun}$ & 3.8 &  5.8    &  7.6    &  7.8     &  8.7    &  9.9    &      \\
                    &  4.28   &  6.15   & \nodata &  8.27    & \nodata & \nodata &     \\
\enddata
\tablenotetext{1}{First Row: Present Work, 25N with mass loss, 25M without mass loss.}
\tablenotetext{2}{Second Row:  Schaller et al. (1992).}
\tablenotetext{3}{Third Row:  LSC (2000).}
\tablenotetext{4}{M$_{bottom}$ is the maximum depth reached by the convective
envelope. Values of Schaller et al. are not available.}
\tablenotetext{5}{$[$Na/Fe$]$ = log($^{23}$Na/Fe)$_{star}$ - log($^{23}$Na/Fe)$_{\sun}$}
\label{tab:SurfAbundRatio}
\end{deluxetable}

\begin{deluxetable}{ccccccccccc}
\tabletypesize{\small}
\rotate
\tablecolumns{11}
\tablewidth{0pc}
\tablecaption{
CHARACTERISTICS OF EVOLUTIONARY MODELS}
\tablehead{ 
\colhead{stage}                  & \colhead{Mass} & \colhead{Evol. Time} & 
\colhead{log(L/L$_{\sun}$)}      & \colhead{log(T$_{\rm eff}$)}          & 
\colhead{T$_{c}^{\; a}$}         & \colhead{$\rho_{\rm c}^{\; b}$}       & 
\colhead{M$_{\alpha}^{\; c}$}    & \colhead{M$_{\rm co}^{\; d}$}         &
\colhead{M$_{\rm ONeMg}^{\; e}$} & \colhead{M$_{\rm cc}^{max \;f}$}     \\
  & \colhead{(M$_{\sun}$)} & \colhead{(yrs)}  & \colhead{} & \colhead{(K)} &
    \colhead{($10^8$ K)} & \colhead{(g cm$^{-3}$)}& \colhead{(M$_{\sun}$)} &
    \colhead{(M$_{\sun}$)} & \colhead{(M$_{\sun}$)} & \colhead{(M$_{\sun}$)} 
}
\startdata
\multicolumn{11}{c}{\underline{15 M$_{\sun}$} }  \\
1 & 15.00 & 0.00                    & 4.277 & 4.487 & 0.342 & 6.28$\times$10$^0$ & 0.00 & 0.00  & 0.00 & 5.66  \\
2 & 14.77 & +1.04$\times$10$^7$     & 4.617 & 4.412 & 0.622 & 6.33$\times$10$^1$ & 2.22 & 0.00  & 0.00 & 0.00  \\
3 & 14.76 & +3.40$\times$10$^4$     & 4.641 & 4.237 & 1.28  & 1.34$\times$10$^3$ & 2.61 & 0.00  & 0.00 & 2.22  \\ 
4 & 13.90 & +1.66$\times$10$^6$     & 4.690 & 3.568 & 3.15  & 6.53$\times$10$^3$ & 3.85 & 1.82  & 0.00 & 0.00  \\
5 & 13.84 & +2.96$\times$10$^4$     & 4.830 & 3.553 & 5.67  & 1.25$\times$10$^5$ & 3.85 & 2.06  & 0.00 & 0.55  \\
6 & 13.82 & +6.76$\times$10$^3$     & 4.859 & 3.550 & 9.95  & 4.87$\times$10$^6$ & 3.85 & 2.18  & 0.44 & 0.00  \\
7 & 13.82 & +8.63$\times$10$^0$     & 4.859 & 3.550 & 12.06 & 8.45$\times$10$^6$ & 3.85 & 2.19  & 0.88 & 0.70  \\
8 & 13.82 & +3.13$\times$10$^0$     & 4.859 & 3.550 & 15.13 & 6.50$\times$10$^6$ & 3.85 & 2.19  & 2.01 & 0.00  \\
9 & 13.82 & +9.82$\times$10$^{-2}$  & 4.859 & 3.550 & 15.01 & 6.73$\times$10$^6$ &
3.85 & 2.19  & 2.01 & 0.00  \\      
10 & 13.82 & +3.71$\times$10$^{0}$  & 4.857 & 3.550 & 19.37 & 2.13$\times$10$^7$ & 3.85 & 2.19  & 2.01  & 0.77    
\\
\\
\\
\multicolumn{11}{c}{\underline{20 M$_{\sun}$} }  \\*
1 & 20.00  & 0.00                   & 4.625 & 4.539 & 0.359  & 4.87$\times$10$^0$ & 0.00  & 0.00  & 0.00  & 8.72 \\*
2 & 19.45  & +7.40$\times$10$^6$    & 4.965 & 4.447 & 0.729  & 5.71$\times$10$^1$ & 3.90  & 0.00  & 0.00  & 0.00 \\*
3 & 19.44  & +1.45$\times$10$^4$    & 4.991 & 4.291 & 1.35   & 6.59$\times$10$^2$ & 4.21  & 0.00  & 0.00  & 3.70 \\*
4 & 17.34  & +1.01$\times$10$^6$    & 5.017 & 3.557 & 3.46   & 5.40$\times$10$^3$ & 5.69  & 3.35  & 0.00  & 0.00 \\*
5 & 17.25  & +1.42$\times$10$^4$    & 5.111 & 3.547 & 5.80   & 4.83$\times$10$^4$ & 5.70  & 3.34  & 0.00  & 0.49 \\*
6 & 17.22  & +3.30$\times$10$^3$    & 5.135 & 3.544 & 10.830 & 2.77$\times$10$^6$ & 5.70  & 3.52  & 0.53  & 0.00 \\*
7 & 17.22  & +1.13$\times$10$^1$    & 5.134 & 3.544 & 12.168 & 6.01$\times$10$^6$ & 5.70  & 3.52  & 0.80  & 0.73 \\*
8 & 17.22  & +1.21$\times$10$^{0}$    & 5.134 & 3.544 & 16.263 & 4.16$\times$10$^6$ & 5.70  & 3.52  & 1.16  & 0.00 \\*
9 & 17.22  & +4.88$\times$10$^{-3}$ & 5.134 & 3.544 & 16.218 & 4.23$\times$10$^6$ & 5.70  & 3.52  & 1.16  & 0.97 \\*
10 & 17.22 & +8.86$\times$10$^{-1}$ & 5.134 & 3.544 & 24.022 & 1.10$\times$10$^7$ & 5.70  & 3.54  & 1.77  & 0.00    
\\
\\
\multicolumn{11}{c}{\underline{25 M$_{\sun}$}  (25C)} \\*
1 & 25.00  &  0.00                  & 4.871 & 4.542 & 0.325  & 2.72$\times$10$^0$ & 0.00  &  0.00 & 0.00  & 12.6 \\*
2 & 23.94  & +6.01$\times$10$^6$    & 5.227 & 4.396 & 1.06   & 1.59$\times$10$^2$ & 5.88  &  0.00 & 0.00  & 0.00 \\*
3 & 23.94  & +4.60$\times$10$^3$    & 5.238 & 4.308 & 1.40   & 4.20$\times$10$^2$ & 6.00  &  0.00 & 0.00  & 5.35 \\*
4 & 18.49  & +7.47$\times$10$^5$    & 5.236 & 3.550 & 3.62   & 4.59$\times$10$^3$ & 7.57  &  4.66 & 0.00  & 0.00 \\*
5 & 18.47  & +9.70$\times$10$^3$    & 5.309 & 3.542 & 6.14   & 3.71$\times$10$^4$ & 7.58  &  4.86 & 0.00  & 0.42 \\*
6 & 18.30  & +1.54$\times$10$^3$    & 5.351 & 3.529 & 11.58  & 2.91$\times$10$^6$ & 7.58  &  5.04 & 0.59  & 0.00 \\*
7 & 18.30  & +5.27$\times$10$^{-1}$ & 5.319 & 3.521 & 12.41  & 3.64$\times$10$^6$ & 7.58  &  5.04 & 0.59  & 0.72 \\*
8 & 18.30  & +5.83$\times$10$^{-1}$ & 5.257 & 3.526 & 17.94  & 3.73$\times$10$^6$ & 7.58  &  5.08 & 1.53  & 0.00 \\*
9 & 18.30  & +9.58$\times$10$^{-3}$ & 5.253 & 3.525 & 18.10  & 4.17$\times$10$^6$ & 7.58  &  5.08 & 1.53  & 1.08 \\*
10 & 18.30 & +1.81$\times$10$^{-1}$ & 5.233 & 3.529 & 24.65  & 1.70$\times$10$^7$ & 7.58  &  5.12 & 1.80  & 0.00     
\\
\\
\multicolumn{11}{c}{\underline{25 M$_{\sun}$}  (25K)} \\*
1 & 25.00  &  0.00                  & 4.876 & 4.575 & 0.371  & 4.06$\times$10$^0$ & 0.00  &  0.00 & 0.00  & 12.8 \\*
2 & 23.95  & +6.00$\times$10$^6$    & 5.223 & 4.397 & 1.06   & 1.61$\times$10$^2$ & 5.88  &  0.00 & 0.00  & 0.00 \\*
3 & 23.94  & +4.40$\times$10$^3$    & 5.237 & 4.311 & 1.39   & 4.12$\times$10$^2$ & 6.00  &  0.00 & 0.00  & 5.31 \\*
4 & 18.60  & +7.49$\times$10$^5$    & 5.231 & 3.550 & 3.62   & 4.62$\times$10$^3$ & 7.53  &  4.67 & 0.00  & 0.00 \\*
5 & 18.47  & +7.90$\times$10$^3$    & 5.280 & 3.545 & 5.51   & 2.19$\times$10$^4$ & 7.54  &  4.79 & 0.00  & 0.47 \\*
6 & 18.41  & +3.50$\times$10$^3$    & 5.322 & 3.540 & 10.98  & 2.03$\times$10$^6$ & 7.54  &  5.01 & 0.45  & 0.00 \\*
7 & 18.41  & +6.93$\times$10$^0$    & 5.329 & 3.541 & 11.90  & 4.10$\times$10$^6$ & 7.54  &  5.01 & 1.09  & 0.75 \\*
8 & 18.41  & +3.28$\times$10$^{0}$ & 5.337 & 3.538 & 16.66  & 4.09$\times$10$^6$ & 7.54  &  5.01 & 1.09  & 0.00 \\*
9 & 18.41  & +0.00$\times$10$^{0}$ & 5.337 & 3.538 & 16.64  & 4.16$\times$10$^6$ & 7.54  &  5.01 & 1.09  & 1.06 \\*
10 & 18.41 & +5.10$\times$10$^{-1}$ & 5.319 & 3.529 & 22.75  & 1.72$\times$10$^7$ & 7.54  &  5.03 & 2.09  & 0.00     
\\
\\
\multicolumn{11}{c}{\underline{25 M$_{\sun}$} (25N)}  \\*
1 & 25.00  &  0.00                  & 4.877 & 4.575 & 0.371 & 4.06$\times$10$^0$ & 0.00  &  0.00 & 0.00  & 12.8 \\*
2 & 23.94  & +5.96$\times$10$^6$    & 5.226 & 4.397 & 1.053 & 1.57$\times$10$^2$ & 5.88  &  0.00 & 0.00  & 0.00 \\*
3 & 23.94  & +4.30$\times$10$^3$    & 5.237 & 4.318 & 1.362 & 3.86$\times$10$^2$ & 6.00  &  0.00 & 0.00  & 5.32 \\*
4 & 18.67  & +7.54$\times$10$^5$    & 5.232 & 3.551 & 3.624 & 4.63$\times$10$^3$ & 7.51  &  4.65 & 0.00  & 0.00 \\*
5 & 18.52  & +9.60$\times$10$^3$    & 5.304 & 3.542 & 6.103 & 3.61$\times$10$^4$ & 7.54  &  4.85 & 0.00  & 0.36 \\*
6 & 18.49  & +1.60$\times$10$^3$    & 5.321 & 3.540 & 12.11 & 1.80$\times$10$^6$ & 7.54  &  5.02 & 0.36  & 0.00 \\*
7 & 18.49  & +1.22$\times$10$^{-1}$ & 5.320 & 3.540 & 12.70 & 2.11$\times$10$^6$ & 7.54  &  5.02 & 0.59  & 0.61 \\*
8 & 18.49  & +2.94$\times$10$^{-1}$ & 5.320 & 3.540 & 17.00 & 2.35$\times$10$^6$ & 7.53  &  5.03 & 1.46  & 0.00 \\*
9 & 18.49  & +9.14$\times$10$^{-4}$ & 5.320 & 3.540 & 17.00 & 2.35$\times$10$^6$ & 7.53  &  5.03 & 1.46  & 0.95 \\*
10 & 18.49 & +3.31$\times$10$^{-1}$ & 5.320 & 3.540 & 22.59 & 9.82$\times$10$^6$ & 7.53  &  5.05 & 1.96  & 0.00  
\\
\\
\\
\multicolumn{11}{c}{\underline{25 M$_{\sun}$} (25NM)}  \\*
1 & 25.00  &  0.00                  & 4.877 & 4.575 & 0.371  & 4.06$\times$10$^0$ & 0.00  &  0.00 & 0.00  & 12.8 \\ *
2 & 25.00  & +5.93$\times$10$^6$    & 5.260 & 4.402 & 1.06   & 1.60$\times$10$^2$ & 6.07  &  0.00 & 0.00  & 0.00 \\*
3 & 25.00  & +4.30$\times$10$^3$    & 5.272 & 4.324 & 1.39   & 4.07$\times$10$^2$ & 6.20  &  0.00 & 0.00  & 5.20 \\*
4 & 25.00  & +7.90$\times$10$^5$    & 5.222 & 3.564 & 3.65   & 4.71$\times$10$^3$ & 7.62  &  4.79 & 0.00  & 0.00 \\*
5 & 25.00  & +1.01$\times$10$^4$    & 5.305 & 3.555 & 6.00   & 3.42$\times$10$^4$ & 7.63  &  4.86 & 0.00  & 0.23 \\*
6 & 25.00  & +1.86$\times$10$^3$    & 5.322 & 3.553 & 11.49  & 1.83$\times$10$^6$ & 7.63  &  4.92 & 0.49  & 0.00 \\*
7 & 25.00  & +1.02$\times$10$^0$    & 5.322 & 3.552 & 12.59  & 2.71$\times$10$^6$ & 7.63  &  4.92 & 0.70  & 0.64 \\*
8 & 25.00  & +3.48$\times$10$^{-1}$ & 5.323 & 3.552 & 17.88  & 3.26$\times$10$^6$ & 7.63  &  4.92 & 1.59  & 0.00 \\*
9 & 25.00  & +4.12$\times$10$^{-3}$ & 5.323 & 3.552 & 17.95  & 3.38$\times$10$^6$ & 7.63  &  4.92 & 1.59  & 1.19 \\*
10 & 25.00 & +1.67$\times$10$^{-1}$ & 5.321 & 3.552 & 25.24  & 5.59$\times$10$^6$ & 7.63  &  4.95 & 2.35  & 0.00   
\\
\\
\multicolumn{11}{c}{\underline{25 M$_{\sun}$}  (25L)} \\*
1 & 25.00  &  0.00                  & 4.873 & 4.572 & 0.371  & 4.06$\times$10$^0$ & 0.00  &  0.00 & 0.00  & 12.8 \\*
2 & 23.95  & +6.02$\times$10$^6$    & 5.166 & 4.453 & 0.79   & 5.28$\times$10$^1$ & 5.61  &  0.00 & 0.00  & 0.00 \\*
3 & 23.95  & +9.90$\times$10$^3$    & 5.167 & 4.222 & 1.40   & 4.60$\times$10$^2$ & 5.75  &  0.00 & 0.00  & 6.16 \\*
4 & 13.86  & +6.88$\times$10$^5$    & 5.336 & 3.522 & 3.56   & 3.94$\times$10$^3$ & 8.34  &  5.53 & 0.00  & 0.00 \\*
5 & 13.66  & +7.70$\times$10$^3$    & 5.382 & 3.522 & 5.99   & 2.66$\times$10$^4$ & 8.36  &  5.61 & 0.00  & 0.47 \\*
6 & 13.61  & +1.88$\times$10$^3$    & 5.382 & 3.522 & 11.18  & 1.94$\times$10$^6$ & 8.36  &  5.69 & 0.48  & 0.00 \\*
7 & 13.61  & +4.35$\times$10$^0$    & 5.382 & 3.522 & 12.05  & 3.63$\times$10$^6$ & 8.36  &  5.72 & 0.74  & 0.77 \\*
8 & 13.61  & +2.30$\times$10$^{0}$  & 5.382 & 3.522 & 17.25  & 3.65$\times$10$^6$ & 8.36  &  5.72 & 1.10  & 0.00 \\*
9 & 13.61  & +8.24$\times$10$^{-3}$ & 5.382 & 3.522 & 17.27  & 3.72$\times$10$^6$ & 8.36  &  5.72 & 1.10  & 1.15 \\*
10 & 13.61 & +4.72$\times$10$^{-1}$ & 5.382 & 3.522 & 22.95  & 1.82$\times$10$^7$ & 8.36  &  5.75 & 2.29  & 0.00   
\\
\\
\\
\\
\multicolumn{11}{c}{\underline{30 M$_{\sun}$} } \\*
1  & 30.00 & 0.00                   & 5.069 & 4.601 & 0.381  & 3.54$\times$10$^0$ & 0.00  &  0.00 & 0.00  & 16.10 \\*
2  & 28.25 & +5.09$\times$10$^6$    & 5.409 & 4.355 & 1.25   & 1.98$\times$10$^2$ & 7.90  &  0.00 & 0.00  & 0.00  \\*
3  & 28.25 & +1.80$\times$10$^3$    & 5.415 & 4.303 & 1.44   & 3.13$\times$10$^2$ & 7.98  &  0.00 & 0.00  & 7.17  \\*
4  & 17.07 & +6.29$\times$10$^5$    & 5.408 & 3.542 & 3.74   & 4.07$\times$10$^3$ & 9.49  & 6.38  & 0.00  & 0.00  \\*
5  & 16.84 & +7.00$\times$10$^3$    & 5.463 & 3.537 & 6.27   & 2.84$\times$10$^4$ & 9.54  & 6.50  & 0.00  & 0.39  \\*
6  & 15.24 & +1.01$\times$10$^3$    & 5.463 & 3.537 & 11.92  & 1.76$\times$10$^6$ & 9.54  & 6.72  & 0.67  & 0.00  \\*
7  & 15.24 & +7.13$\times$10$^{-1}$ & 5.463 & 3.537 & 12.44  & 2.37$\times$10$^6$ & 9.54  & 6.75  & 0.80  & 0.67  \\*
8  & 15.24 & +3.34$\times$10$^{-1}$ & 5.463 & 3.537 & 18.70  & 3.06$\times$10$^6$ & 9.54  & 6.75  & 1.69  & 0.00  \\*
9  & 15.24 & +4.58$\times$10$^{-3}$ & 5.463 & 3.537 & 18.21  & 2.68$\times$10$^6$ & 9.54  & 6.75  & 1.78  & 1.20  \\*
10 & 15.24 & +1.02$\times$10$^{-1}$ & 5.463 & 3.537 & 26.02  & 4.68$\times$10$^6$ & 9.54  & 6.89  & 2.80  & 0.00     
\\
\\
\\
\\
\\
     \multicolumn{11}{c}{} \\*
   \nodata \\*
   \nodata \\*
   \nodata \\*
\enddata
\tablenotetext{a}{T$_{C}$ is the central temperature.}
\tablenotetext{b}{$\rho_{C}$ is the central density.}
\tablenotetext{c}{M$_{\alpha}$ is the mass size of helium core.}
\tablenotetext{d}{M$_{co}$ is the mass size of C/O core.}
\tablenotetext{e}{M$_{ONeMg}$ is the mass size of O/Ne/Mg core.}
\tablenotetext{f}{M$_{cc}^{max}$ is the maximum convective core mass of
that burning phase.}
\tablenotetext{*}{Stages 1 to 10 correspond to the evolutionary time of
1) Zero Age Main sequence, 2) central hydrogen exhaustion, 
3) central helium ignition, 4) central helium exhaustion, 
5) central carbon ignition, 6) central carbon exhaustion,
and 7) central neon ignition, 8) central neon exhaustion,
9) central oxygen ignition, 10) central oxygen exhaustion,
respectively.}
\label{tab:evolmod}
\end{deluxetable}
 \clearpage

\begin{deluxetable}{lcccccc}
\footnotesize
\tabletypesize{\small}
\rotate
\tablecolumns{7}
\tablecaption{Central $^{12}$C and 
$^{20}$Ne Mass Fractions at Central Core Helium and 
at Central Core Neon Exhaustions,
$^{16}$O Mass Fractions at 
Central Core Helium,
at Central Core Carbon, and at Central Core Neon Exhaustions}
\tablewidth{0pc}
\tablehead{
\colhead{Mass} & \colhead{\underline{$X(^{12}{\rm C})$}}
               & \multicolumn{2}{c}{\underline{$X(^{20}{\rm Ne})$}}  
               & \multicolumn{3}{c}{\underline{$X(^{16}{\rm O})$}} \\
($\msun$)     & End Helium & End Helium & End Carbon & End Helium & End Carbon & End Neon 
}
\startdata
15   & 0.285 & 0.0028 & 0.371 & 0.688 & 0.548 & 0.738  \\
20   & 0.258 & 0.0041 & 0.349 & 0.713 & 0.579 & 0.758  \\
25K  & 0.280 & 0.0052 & 0.384 & 0.689 & 0.541 & 0.737  \\
25L  & 0.274 & 0.0054 & 0.379 & 0.695 & 0.549 & 0.740  \\
25N  & 0.236 & 0.0052 & 0.329 & 0.733 & 0.606 & 0.795  \\
25NM & 0.193 & 0.0067 & 0.275 & 0.774 & 0.669 & 0.806  \\   
30   & 0.221 & 0.0070 & 0.316 & 0.745 & 0.624 & 0.786  \\
\enddata
\label{tab:COMassfrac}
\end{deluxetable}
\clearpage

\clearpage

\begin{deluxetable}{lccc}
\footnotesize
\tablecolumns{4}
\tablecaption{Comparison of 15 M$_{\sun}$ Models}
\tablewidth{0pc}
\tablehead{
\colhead{Physical Quantity}     & \colhead{PW$^a$} & \colhead{LSC$^b$} & \colhead{WHW$^c$}  \\
}
\startdata
 & \multicolumn{3}{c}{H-Burning}  \\
\cline{2-4}
$\tau_H$(10$^6$ yr)               &  10.4       &  10.7       &  11.1    \\
M$_{cc}$ (M$_{\sun}$)             &  5.66       &  6.11       &  5.7     \\
M$_{C-envelope}^{final}$          &  3.94       &  4.28       &  4.5     \\
M$_{core}^*$                      &  3.85       &  4.10       &  4.15    \\
\hline
\\

 & \multicolumn{3}{c}{He-Burning}  \\
\cline{2-4}
$\tau_{He}$ (10$^6$ yr)           &  1.66       &  1.40       &  1.97    \\
M$_{cc}$ (M$_{\sun}$)             &  2.22       &  2.33       &  2.60    \\
M$_{core}$ (M$_{\sun}$)           &  2.27       & 2.39        &  2.80    \\
X($^{12}$C)/X($^{16}$O)           & 0.285/0.688 & 0.195/0.777 & 0.216/-  \\
$\Delta$M (He-shell) (M$_{\sun}$) & 2.3-3.1     & 2.43-3.54   & 2.8-4.2  \\ 
\hline
\\

 & \multicolumn{3}{c}{C-Burning}  \\*
\cline{2-4}
$\tau_{C}$ (10$^3$ yr)            &  6.76     &  2.60       &  2.03    \\*
M$_{cc}$ (M$_{\sun}$)             &  0.55     &  0.41       &  0.50    \\*
M$_{core}$ (M$_{\sun}$)           &  2.01     &  2.44       &  1.85    \\*
$\Delta$M (C-shell) (M$_{\sun}$)  & 0.50-1.21 & 0.39-0.80   & 0.59-1.20 \\*
                                  & 0.99-1.76 & 0.80-1.18   & 1.20-2.30 \\*
                                  & 1.38-1.91 & 1.20-1.77   & 1.90-2.61  \\*
                                  &           & 1.56-1.67   &           \\*
                                  &           & 1.64-2.18   &           \\* 
\hline
\\

 & \multicolumn{3}{c}{Ne-Burning} \\*
\cline{2-4}
$\tau_{Ne}$ (yr)                  &  1.98      &  2.00       &  0.73    \\*
M$_{cc}$ (M$_{\sun}$)             &  0.70      &  0.66       &  0.70    \\*
X($^{16}$O)/X($^{24}$Mg)          &  0.74/0.09 &  0.81/0.05  & --       \\*
\hline
\\

 & \multicolumn{3}{c}{O-Burning}  \\*
\cline{2-4}
$\tau_{O}$ (yr)                   &  3.71     &  2.47       &  2.58    \\*
M$_{cc}$ (M$_{\sun}$)             &  0.77     &  0.94       &  0.80    \\*
\enddata
\tablenotetext{*}{M$_{core}$ means in this and the following Table the 
region in which the fuel of the burning phase has been exhausted.}
\label{tab:compmod15}
\end{deluxetable}

\clearpage

\clearpage

\begin{deluxetable}{lcccccc}
\tabletypesize{\small}
\tablecolumns{7}
\tablecaption{Comparison of 25 M$_{\sun}$ Models}
\tablewidth{0pc}
\tablehead{
\colhead{Physical Quantity}       & \colhead{25K} & \colhead{25N}  & \colhead{25NM} & \colhead{LSC} & \colhead{25L} & \colhead{WHW}    \\
}
\startdata
\multicolumn{7}{c}{H-Burning}  \\
\cline{2-7}
$\tau_H$(10$^6$ yr)               &  6.00         &   6.00         &   5.93         &   5.93        &  6.02         &  6.70           \\
M$_{cc}$ (M$_{\sun}$)             &  12.8         &   12.8         &   12.8         &   12.8        &  12.8         &  12.5           \\
M$_{core}$                       &  7.54          &   7.53         &   7.63         &   8.01        &  8.36         &  8.43            \\
M$_{conv.-env.}^{final}$          &  7.60         &   7.60         &   7.89         &   8.27        &  8.71         &  8.60           \\
\hline
\\
\multicolumn{7}{c}{He-Burning}  \\
\cline{2-7}
$\tau_{He}$ (10$^6$ yr)           &  0.749        &   0.754        &   0.79         &  0.68         &  0.688        &  0.839         \\
M$_{cc}$ (M$_{\sun}$)             &  5.31         &   5.32         &   5.20         &  5.23         &  6.16         &  6.50          \\
M$_{C/O-core}$ (M$_{\sun}$)       &  5.03         &   5.05         &   4.95         &  4.90         &  6.09         &  6.45           \\
X($^{12}$C)/X($^{16}$O)           & 0.28/0.69     &  0.24/0.73     & 0.19/0.77      &  0.18/0.79    &  0.27/0.70    &  0.19/-         \\
$\Delta$M (He-shell) (M$_{\sun}$) & 5.30-6.30     &  5.30-6.30     & 5.25-7.35      & 5.30-7.68$^a$ &  6.28-8.07    &  6.5-8.1        \\ 
\hline
\\
\multicolumn{7}{c}{C-Burning}  \\*
\cline{2-7}
$\tau_{C}$ (yr)                   &  3500         &  1600          &  1860          &  970          &  1880         &  522            \\*
M$_{cc}$ (M$_{\sun}$)             &  0.47         &   0.36         &  0.23          &  radiative    &  0.47         &  radiative       \\*
M$_{core}^*$ (M$_{\sun}$)         &  2.50         &   1.96         &  2.63          &  2.40         &  2.61         &  2.45            \\*
$\Delta$M (C-shell) (M$_{\sun}$)  & 0.50-1.19     &  0.38-0.97     &  0.865-1.16    &  1.48-2.43    &  0.52-1.30    &  1.60-5.60       \\*
                                  & 1.30-4.54     &  1.04-2.19     &  1.21 - 3.12   &  2.28-4.61    &  1.33-5.15    &  3.20-5.70        \\*
                                  &               &  2.26-4.94     &  2.76 - 3.50   &               &               &                   \\*
\hline
\\
\multicolumn{7}{c}{Ne-Burning} \\*
\cline{2-7}
$\tau_{Ne}$ (yr)                  &  3.28         &   0.294        &   0.348        & 0.77          &   4.35        &  0.89          \\*
M$_{cc}$ (M$_{\sun}$)             &  0.75         &   0.61         &   0.64         & 0.50          &   0.77        &  1.0           \\* 
M$_{core}$ (M$_{\sun}$)           &  1.80         &   1.96         &   2.16         & 2.29          &   2.05        &  2.20          \\*
X($^{16}$O)/X($^{24}$Mg)          &  0.74/0.10    &   0.80/0.09    &   0.81/0.07    & 0.83/0.05     &   0.74/0.10   &   -            \\*
\hline
\\
\multicolumn{7}{c}{O-Burning}  \\
\cline{2-7}
$\tau_{O}$ (yr)                   &  0.510        &   0.33         &   0.547        &  0.33         &   0.47        &  0.40          \\
M$_{cc}$ (M$_{\sun}$)             &  1.06         &   0.95         &   1.19         &  1.15         &   1.15        &  1.25          \\
\enddata
\tablenotetext{*}{The definition of M$_{core}$ is somewhat ambiguous, one should see the carbon profile for more detail.}
\tablenotetext{a}{We took the second broad convective He-shell obtained by Limongi et al. (2000) which resembles ours.}
\label{tab:compmod25}
\end{deluxetable}

\clearpage


~


\clearpage
\setcounter{figure}{0}
\begin{figure}
\epsscale{0.7}
\rotatebox{90}{\plotone{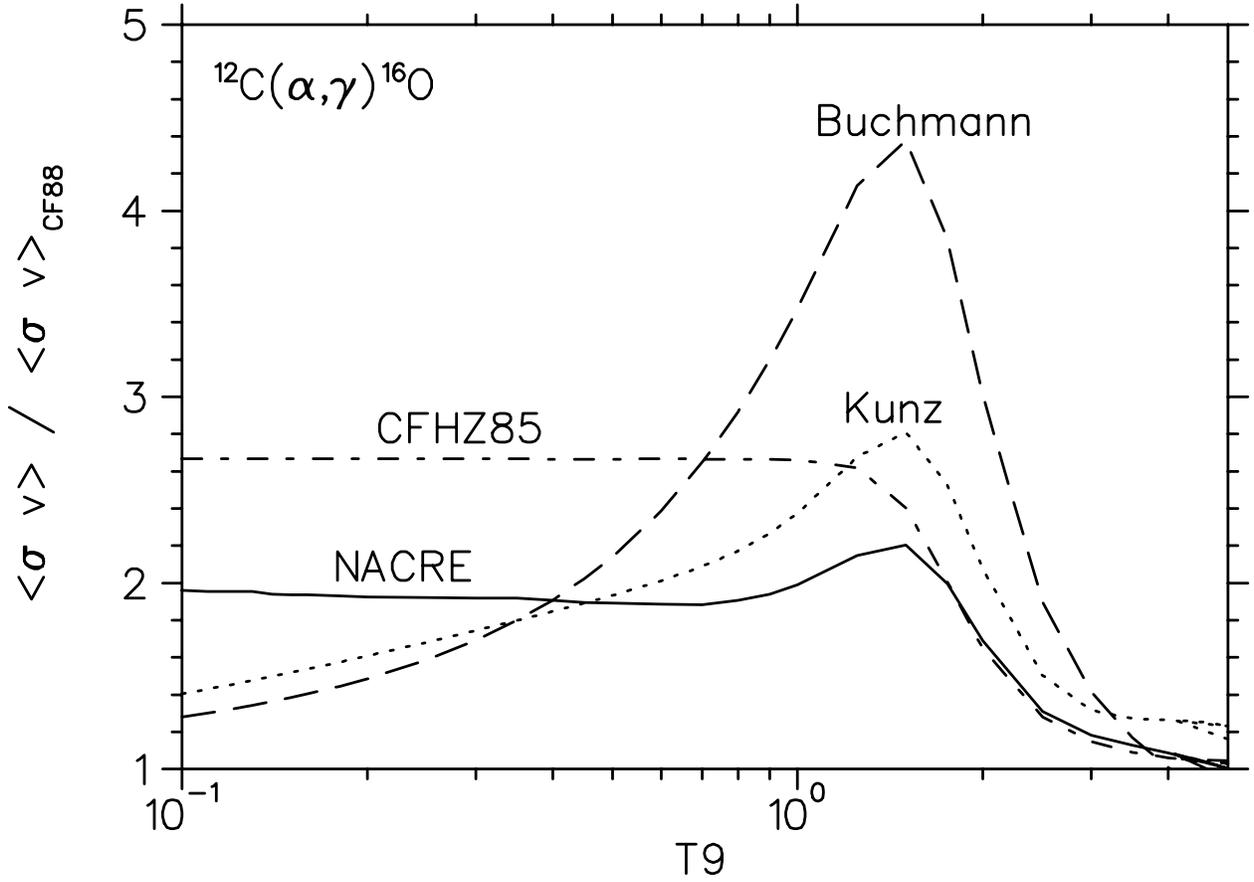}}
\caption{
Comparison of the
\protect{$^{12}$C($\alpha$,$\gamma$)$^{16}$O}
 reaction rates among different
compilations. See text for references.
Note that core helium burning occurs in the range T$_9$ = 0.13 - 0.40
for the stars under investigations in this paper
(see Table \ref{tab:evolmod}).
\label{fig:C12agO16rate}
}
\end{figure}
\clearpage

\begin{figure}
\figurenum{2}
\epsscale{0.7}
\rotatebox{90}{\plottwo{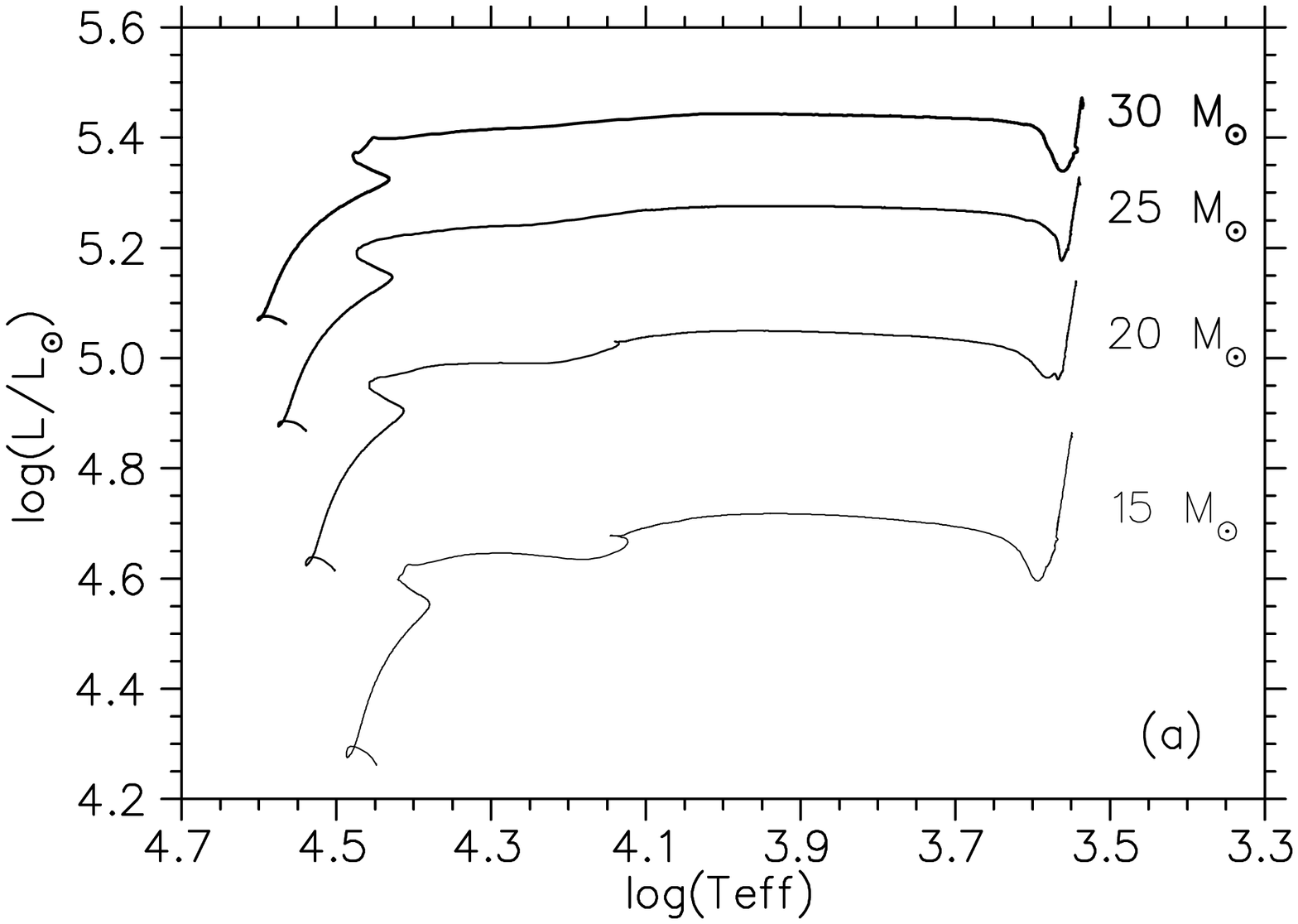}{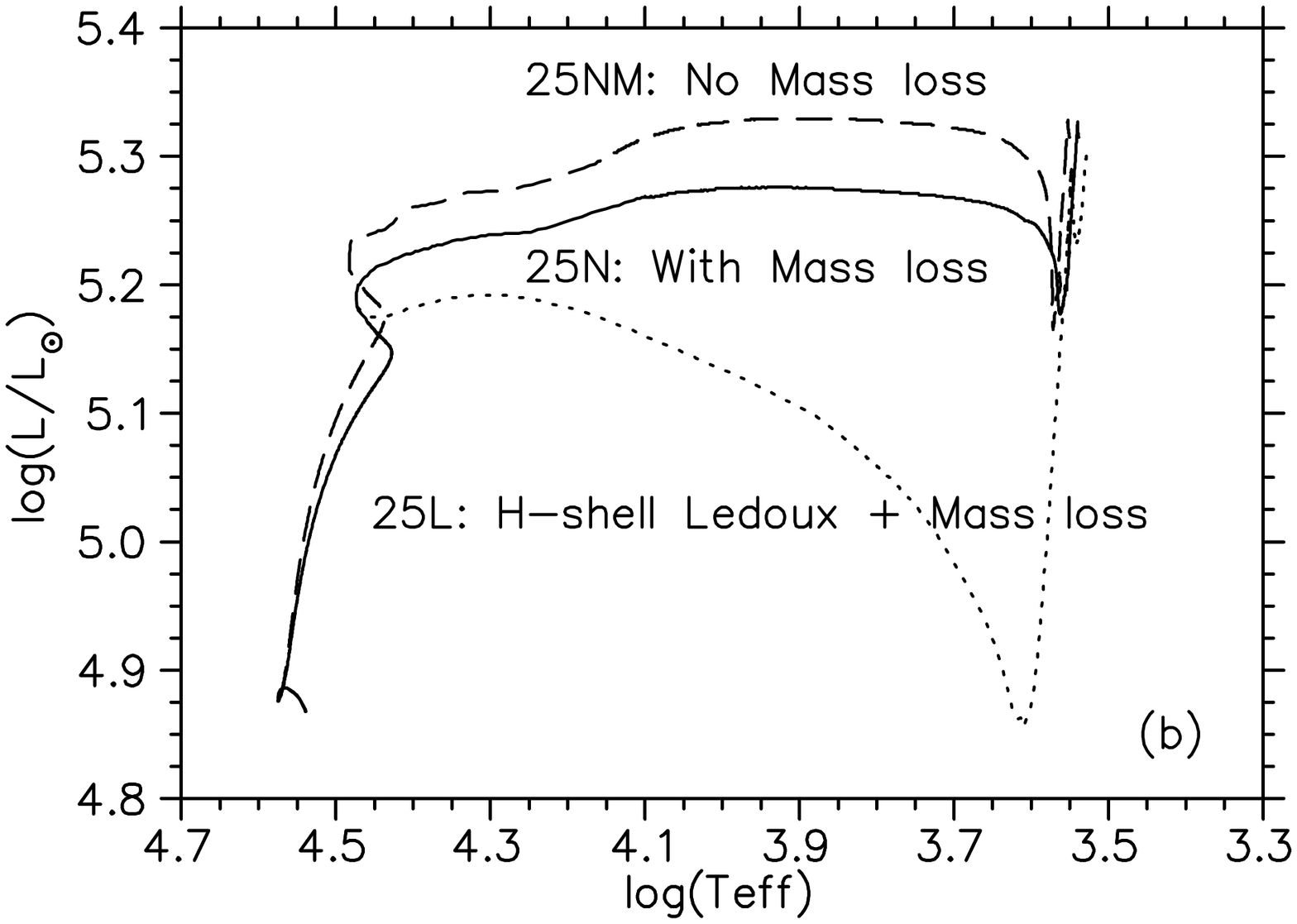}}
\caption{
Hertzsprung-Russell diagram for the stars studied
in this work (a). See Table \ref{tab:stellarModels} for
more information about these tracks.
The Hertzsprung-Russell diagram for the 25 $\msun$ star 
evolved with different assumptions (b).
A significant difference between these tracks is clearly seen.
\label{fig:HRdiagram}
}
\end{figure}
\clearpage

\clearpage

\setcounter{figure}{2}

\begin{figure}
\epsscale{0.7}
\rotatebox{90}{\plotone{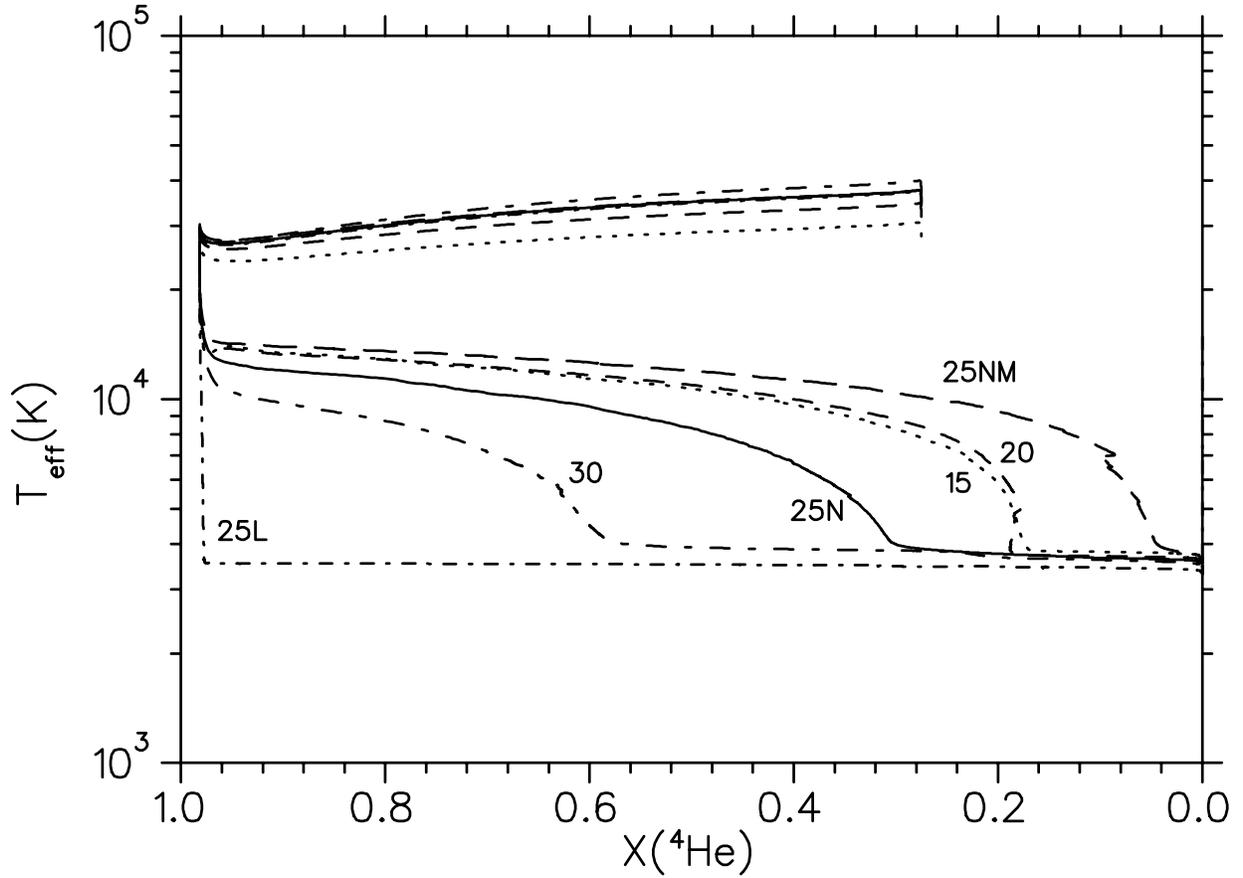}}
\caption{
Evolution of the stars under consideration
in a T$_{eff}$ - X($^4$He) plane, where
X($^4$He) is the central mass fraction of helium.
Note the different timing of the transition to the red giant branch
during core He-burning in particular the immediate evolution of
25L to that stage.
\label{fig:Teff_vs_XHe4}
}
\end{figure}
\clearpage

\setcounter{figure}{3}

\begin{figure}
\epsscale{0.7}
\plotone{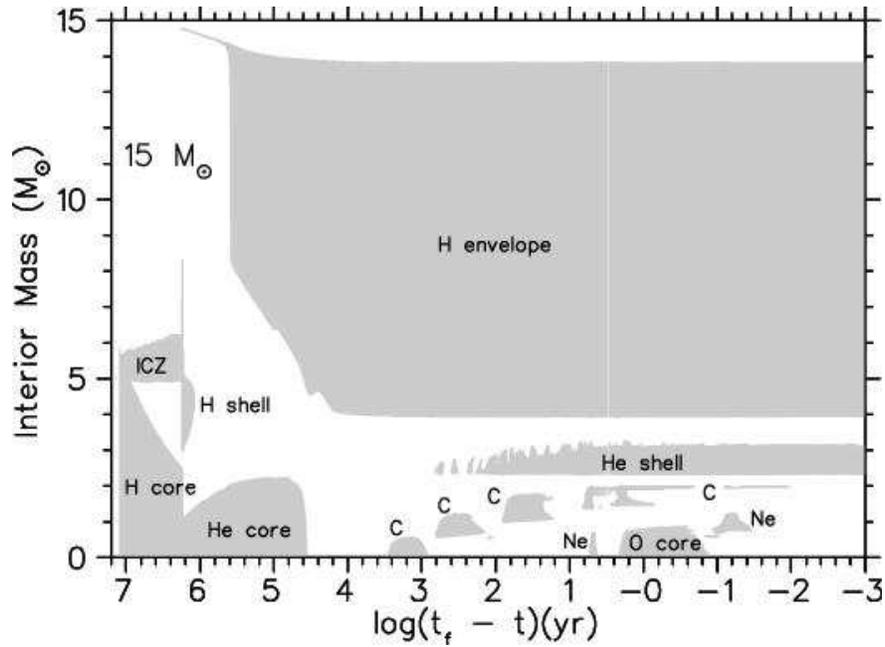}
\caption{
The change of the internal structure as a function of time of 
the 15 $\msun$ star of initial solar-like composition.
The star is evolved with mass loss by stellar wind from the
main sequence till oxygen shell-burning.
Each vertical line corresponds to a stellar model
at a given time. Convective regions are labeled according
to their physical origin, see text for details. Note that 
the t$_f$ is the time of the last stellar calculated and the 
t is the evolution time. For example core hydrogen burning 
in this diagram starts at log(t$_f$ - t)=7.02, or 10$^{7.02}$
years from the time of the last model. 
\label{fig:convective_zones_f15n0}
}
\end{figure}
\clearpage

\setcounter{figure}{4}

\begin{figure}
\epsscale{0.7}
\plotone{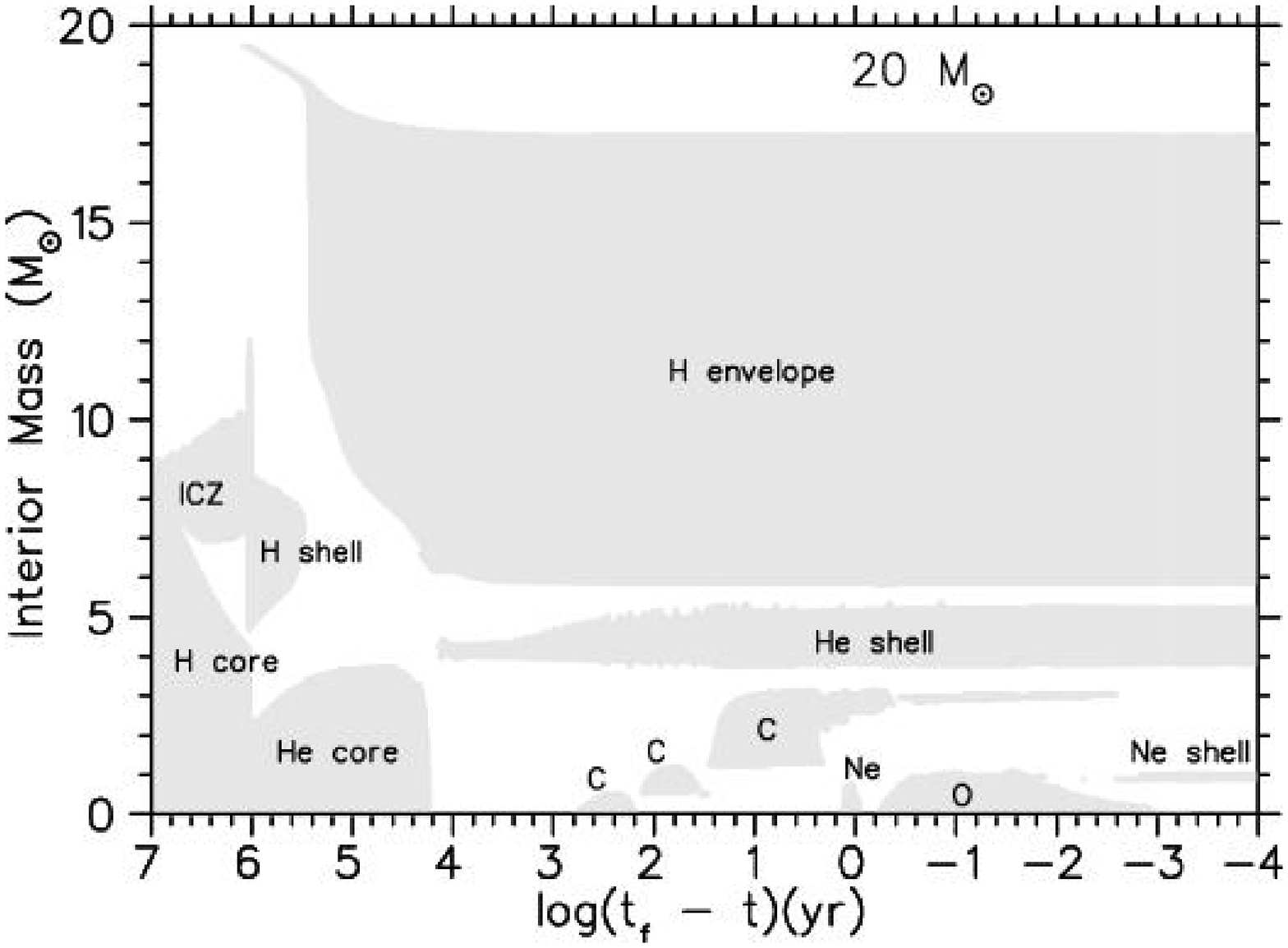}
\caption{
Same as Fig. \ref{fig:convective_zones_f15n0} for the 20 $\msun$ star.
\label{fig:convective_zones_f20n0}
}
\end{figure}
\clearpage

\begin{figure}
\epsscale{0.7}
\plotone{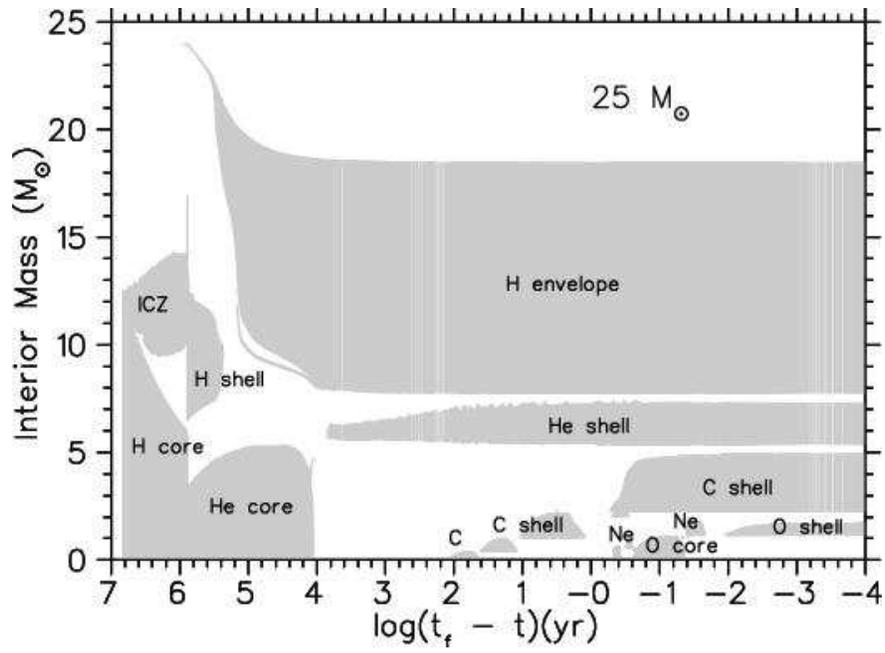}
\caption{
Same as Fig. \ref{fig:convective_zones_f15n0} for a 25 $\msun$ star 
(sequence 25N), 
(see Table \ref{tab:stellarModels} for more information).
Note that the convective zone above the H-burning shell dies out before the
end of core He-burning in contrast to Fig.
\ref{fig:convective_zones_25NM}
without mass loss.
\label{fig:convective_zones_f25n0}
}
\end{figure}
\clearpage

\begin{figure}
\epsscale{0.7}
\plotone{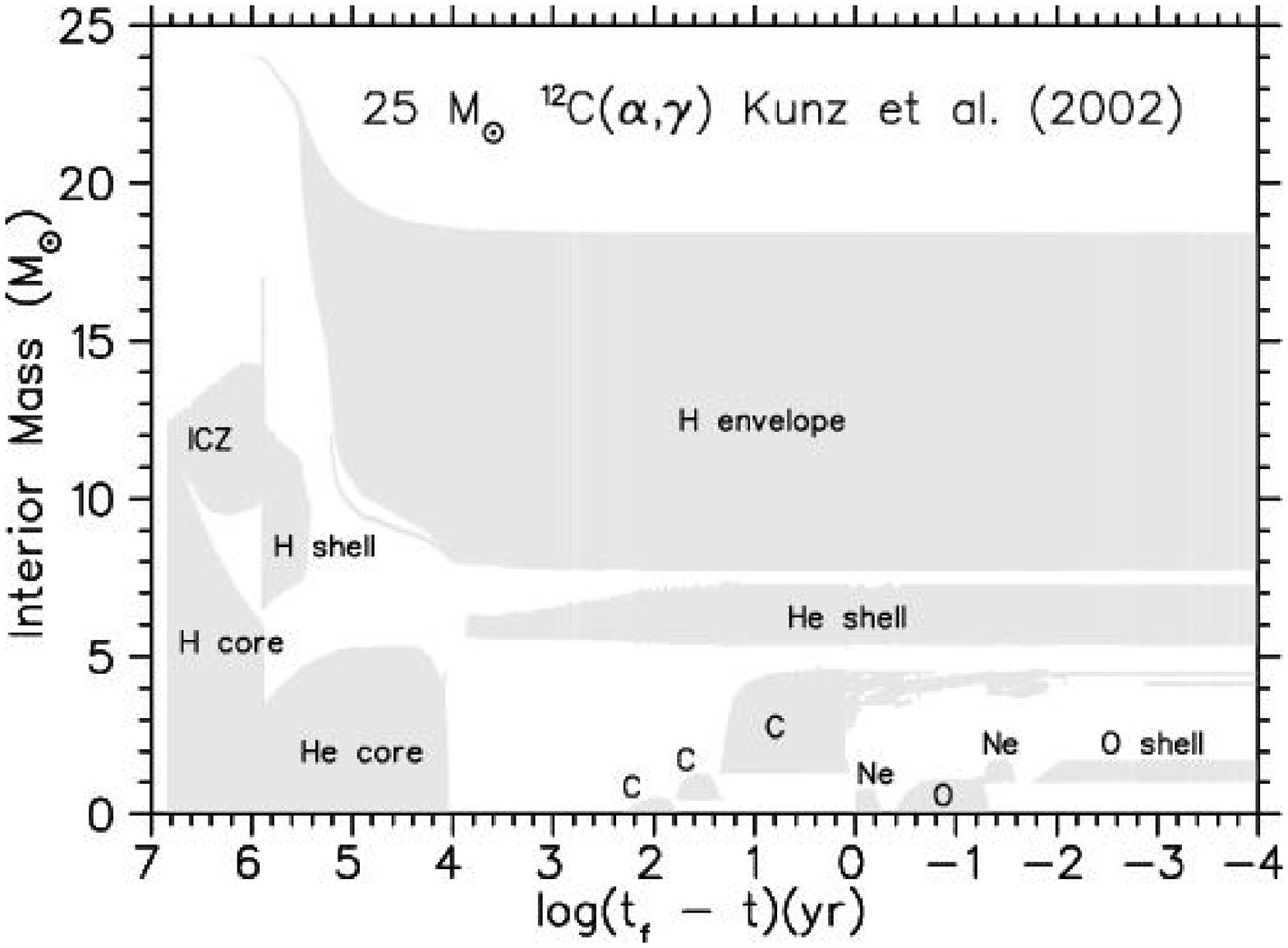}
\caption{
Same as Fig. \ref{fig:convective_zones_f15n0} for the sequence 25K
(See Table \ref{tab:stellarModels} for more information).
\label{fig:convective_zones_f25l0}
}
\end{figure}
\clearpage

\begin{figure}
\epsscale{0.7}
\plotone{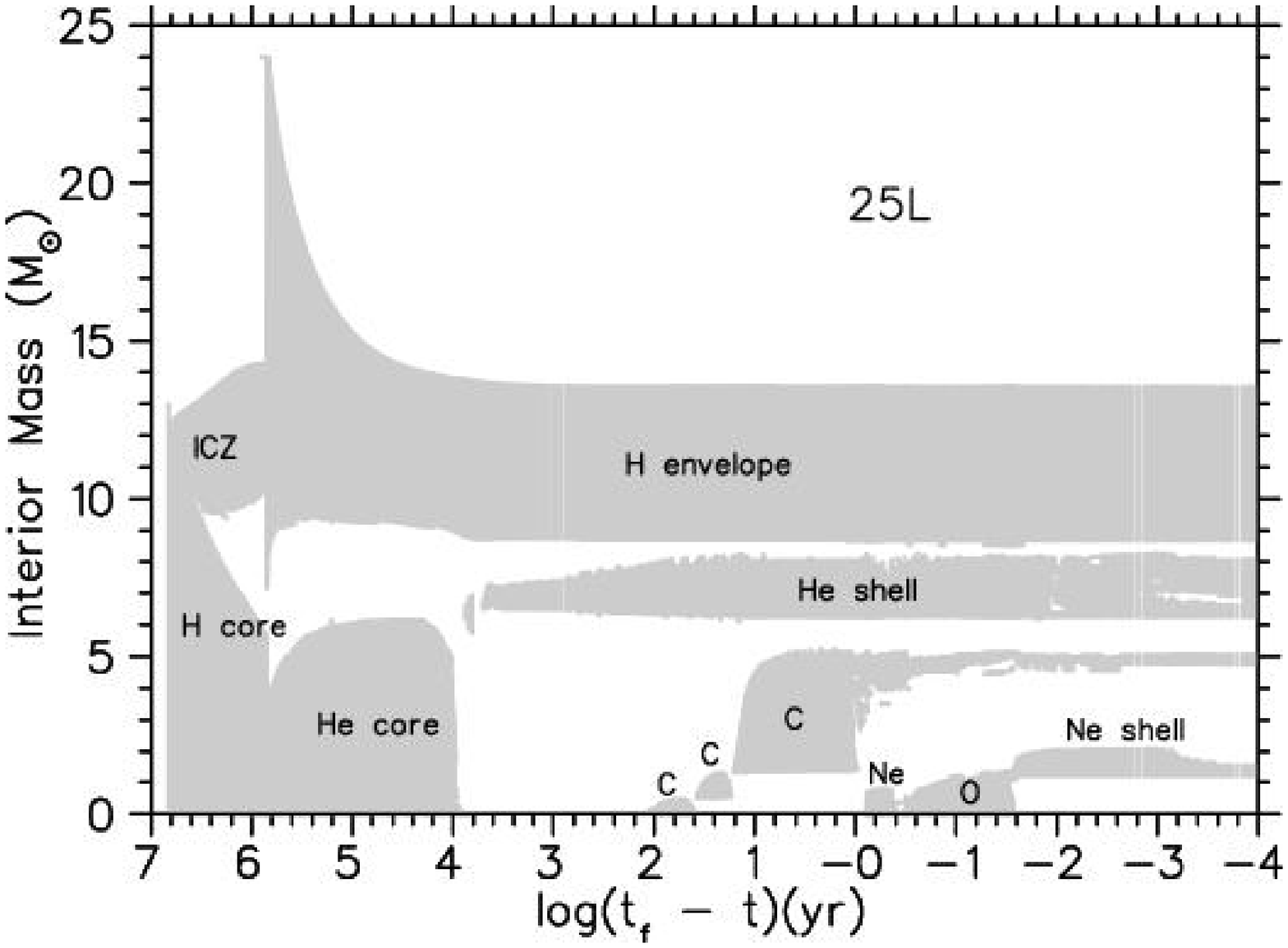}
\caption{
Same as Fig. \ref{fig:convective_zones_f15n0} for the sequence 25L
(See Table \ref{tab:stellarModels} for more information).
\label{fig:convective_zones_f25dx}
}
\end{figure}
\clearpage

\begin{figure}
\epsscale{0.7}
\plotone{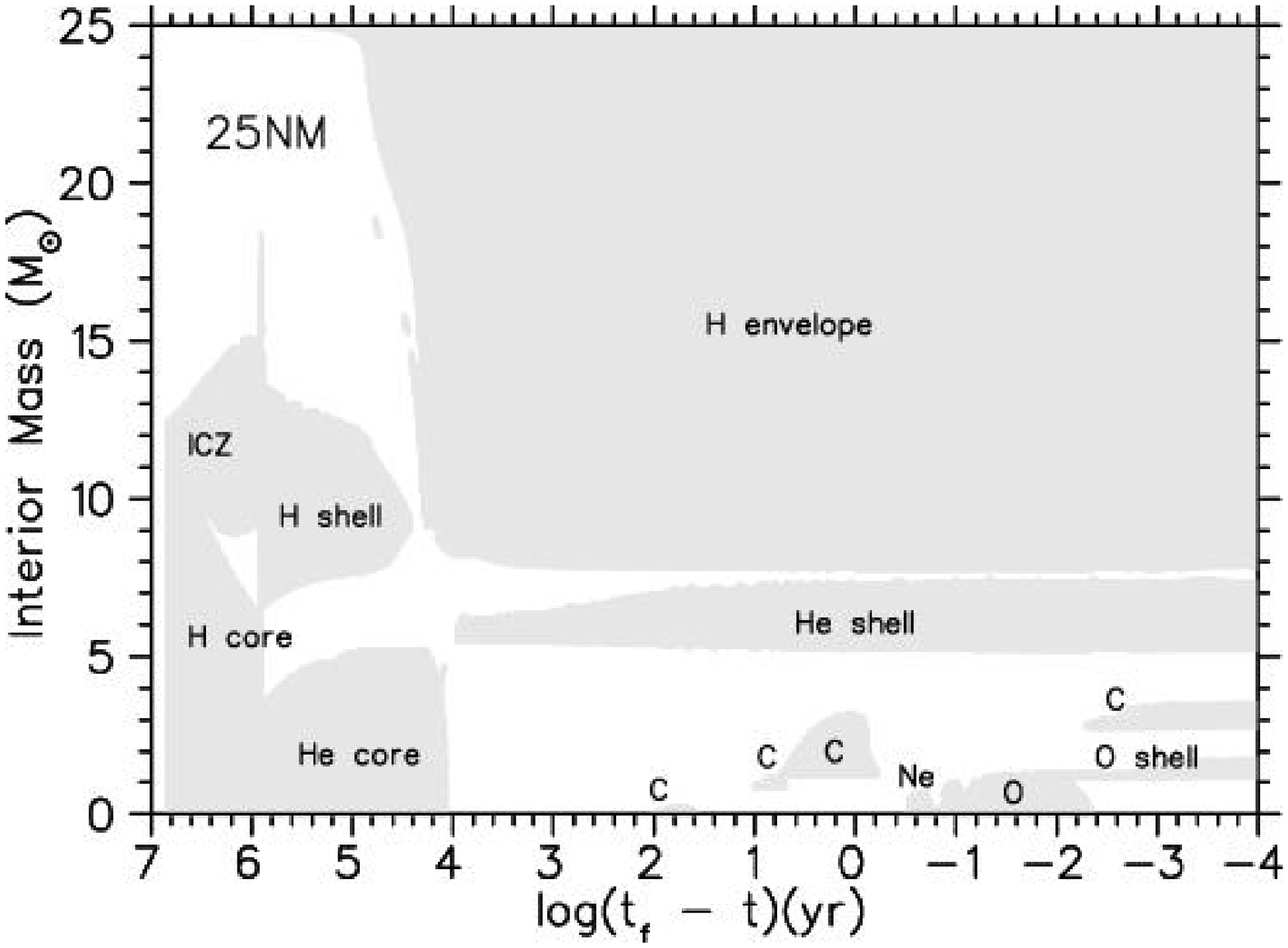}
\caption{
Same as Fig. \ref{fig:convective_zones_f15n0} for the sequence 25NM
(See Table \ref{tab:stellarModels} for more information).
\label{fig:convective_zones_25NM}
}
\end{figure}
\clearpage

\begin{figure}
\epsscale{0.7}
\plotone{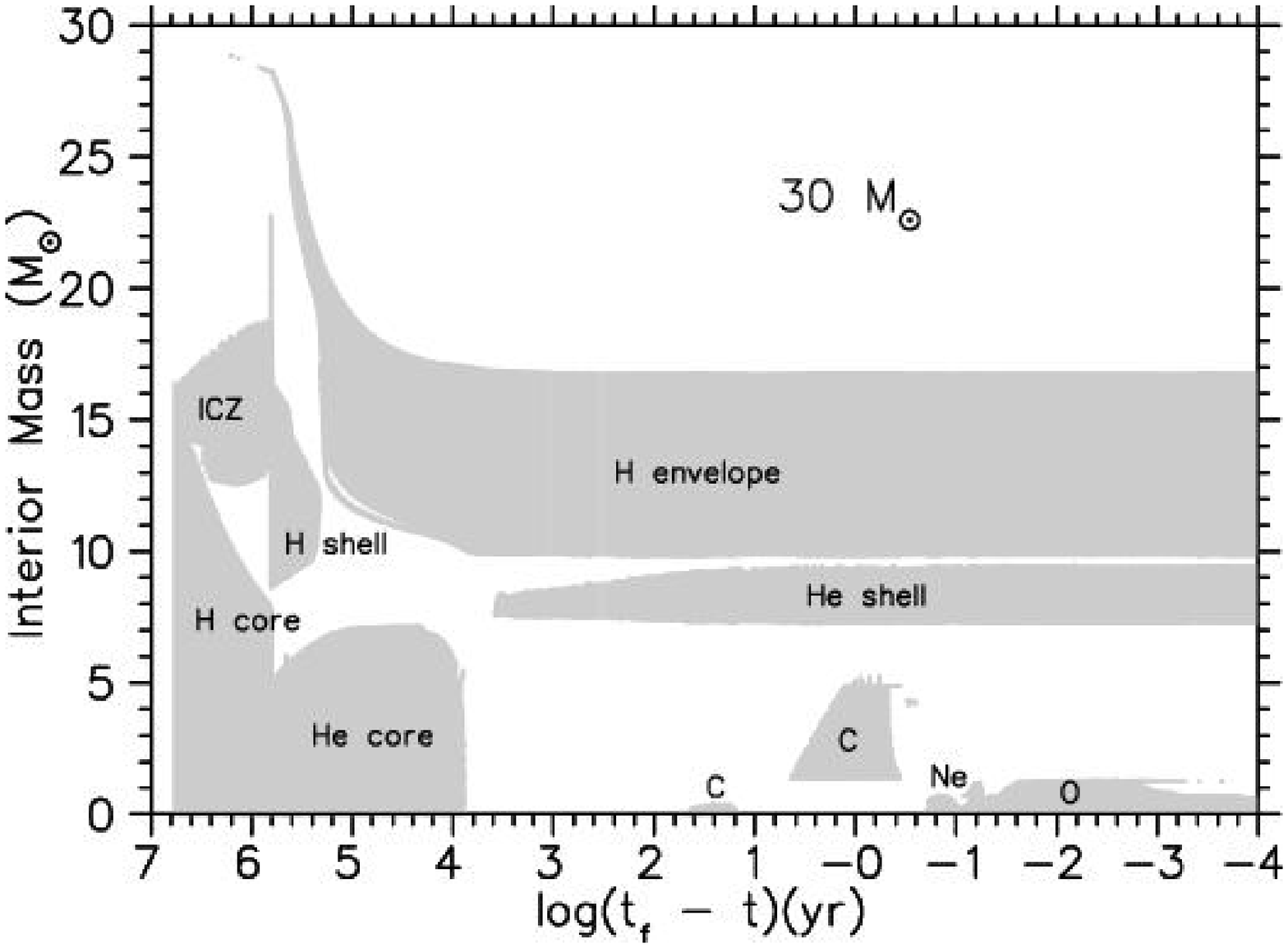}
\caption{
Same as Fig. \ref{fig:convective_zones_f15n0} for the star of 30 $\msun$.
\label{fig:convective_zones_f30n0}
}
\end{figure}
\clearpage

\begin{figure}
\figurenum{11}
\epsscale{0.7}
\rotatebox{90}{\plotone{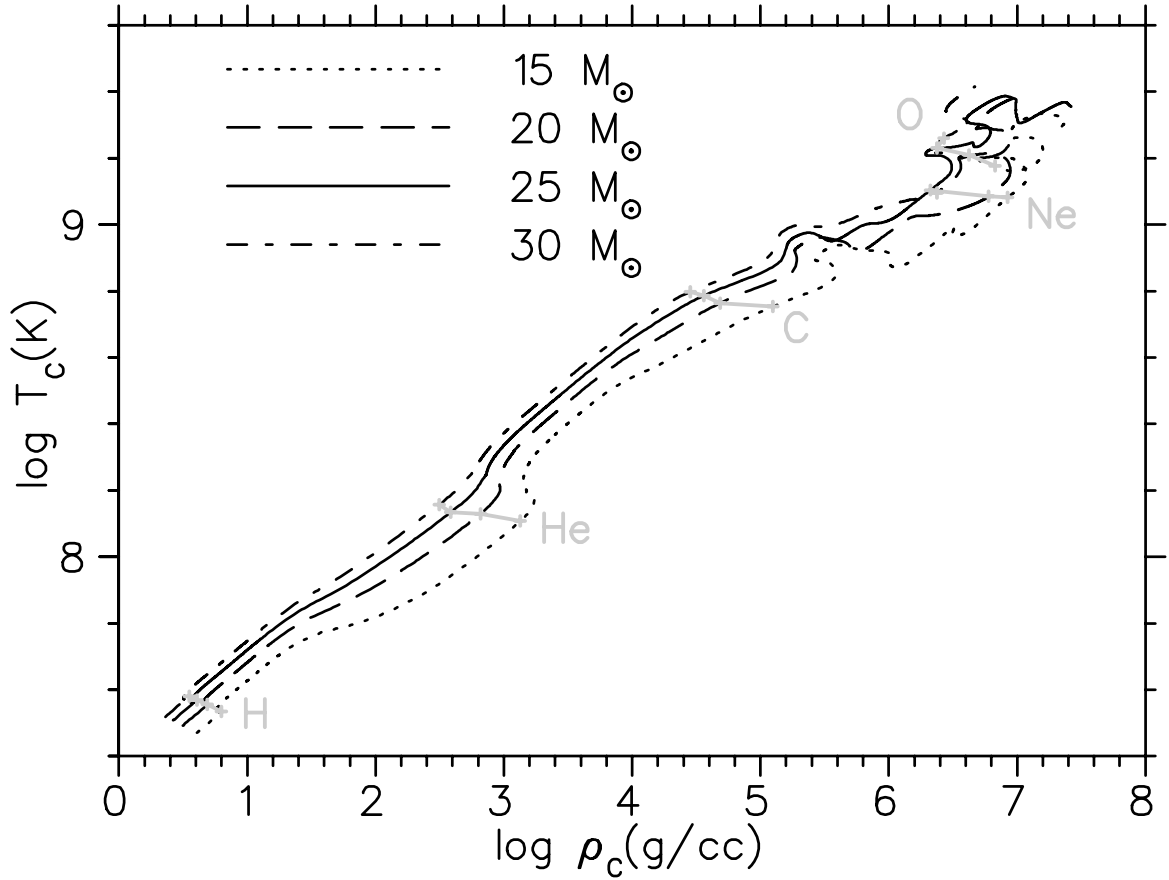}}
\caption{
Central Temperature T$_C$ versus central density $\rho_C$
for the stars of masses 15 $\msun$ to 30 $\msun$.
The lines marking H, He, C, Ne, and O
indicate at which $T_C$ and $\rho_C$
a burning phase starts.
\label{fig:Tc_vs_rhoC}
}
\end{figure}
\clearpage

\begin{figure}
\figurenum{12}
\epsscale{0.7}
\rotatebox{90}{\plottwo{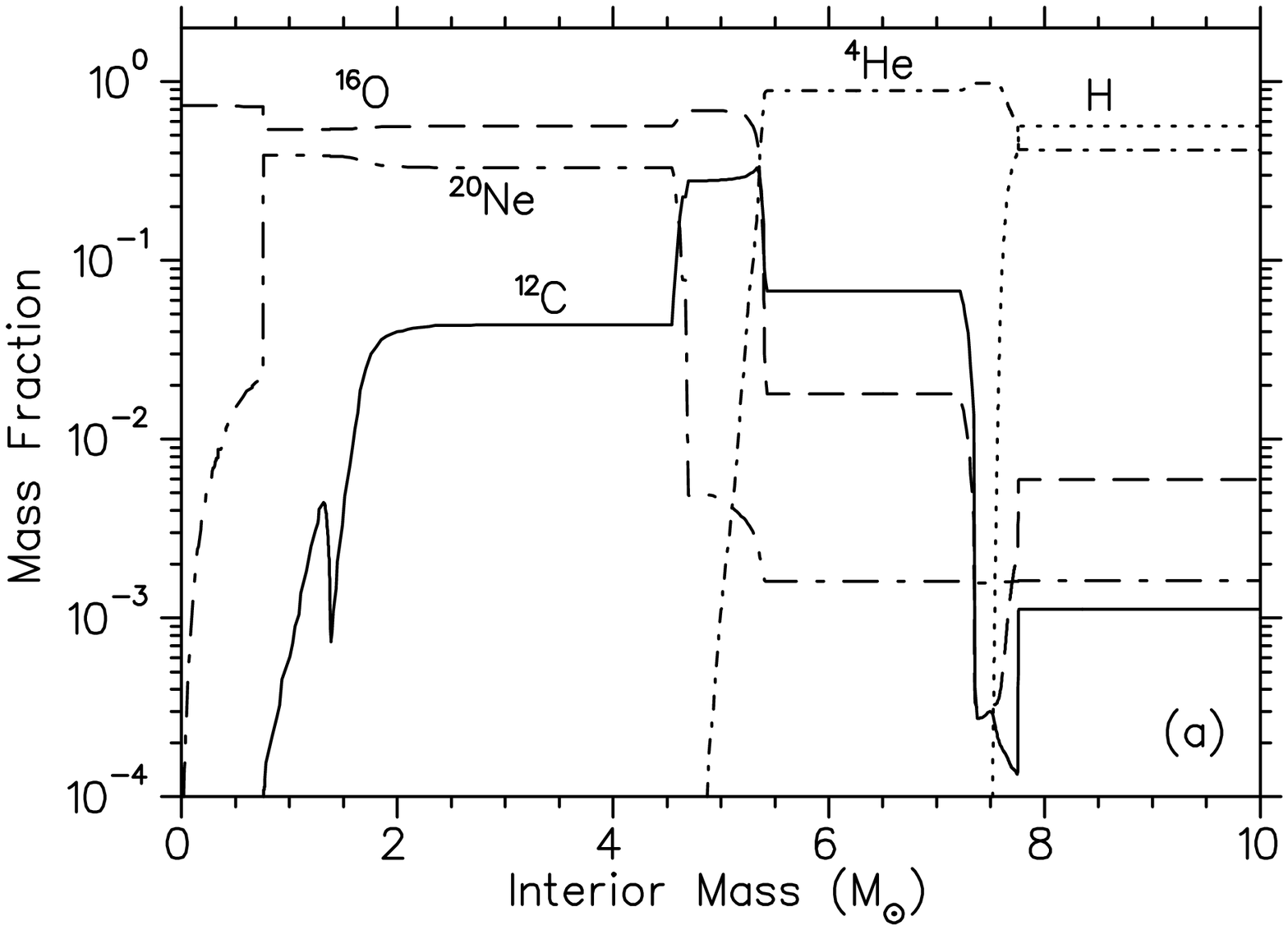}{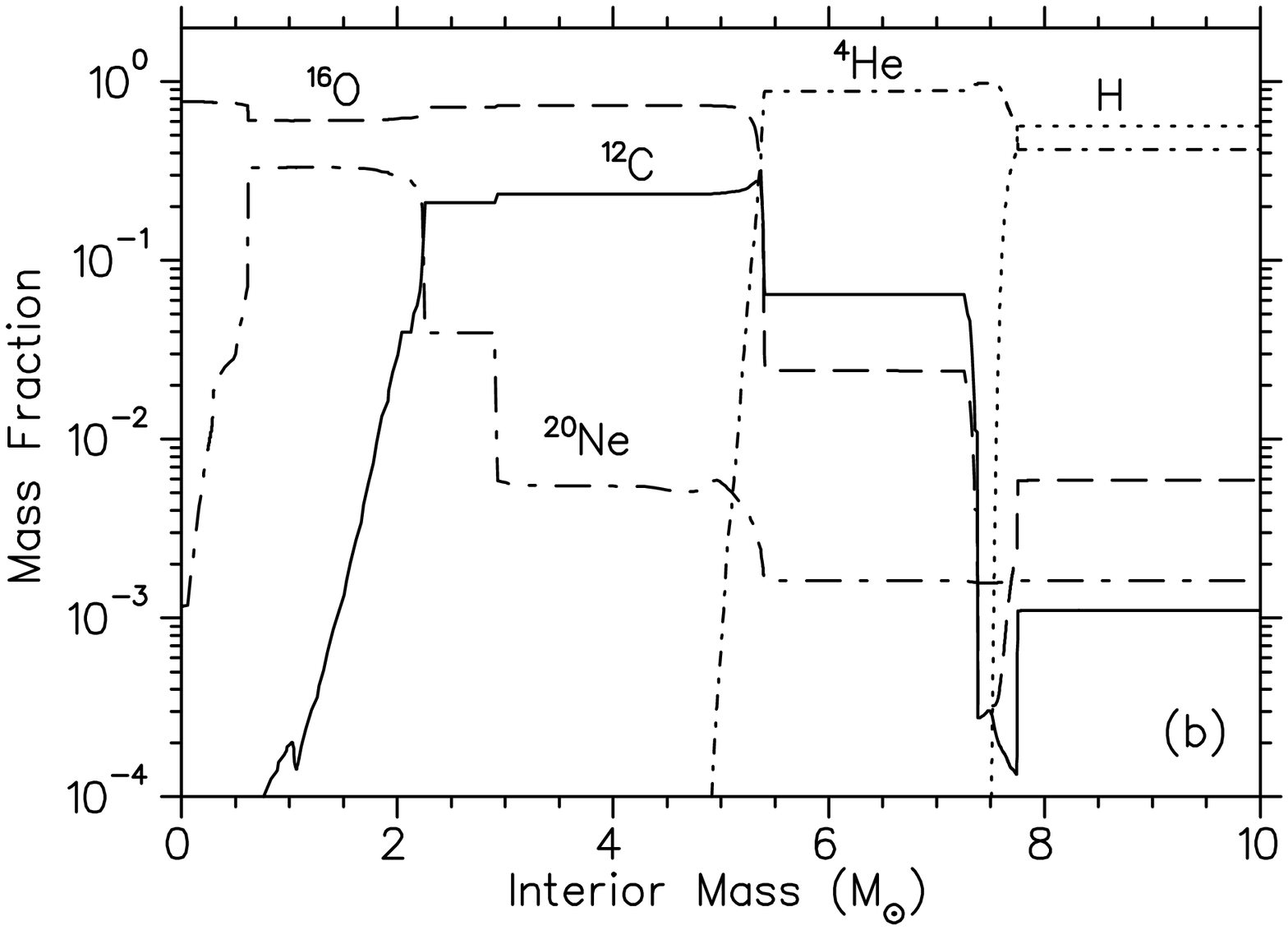}}
\caption{
Composition profile for the sequence 25K (a) and 
the sequence 25N (b) versus
interior mass at the end of core Ne-burning.
\label{fig:s25k0_massfrac_vs_Mr_mod05050_EndofNeburn}
}
\end{figure}
\clearpage

\begin{figure}
\figurenum{13}
\epsscale{0.7}
\rotatebox{90}{\plottwo{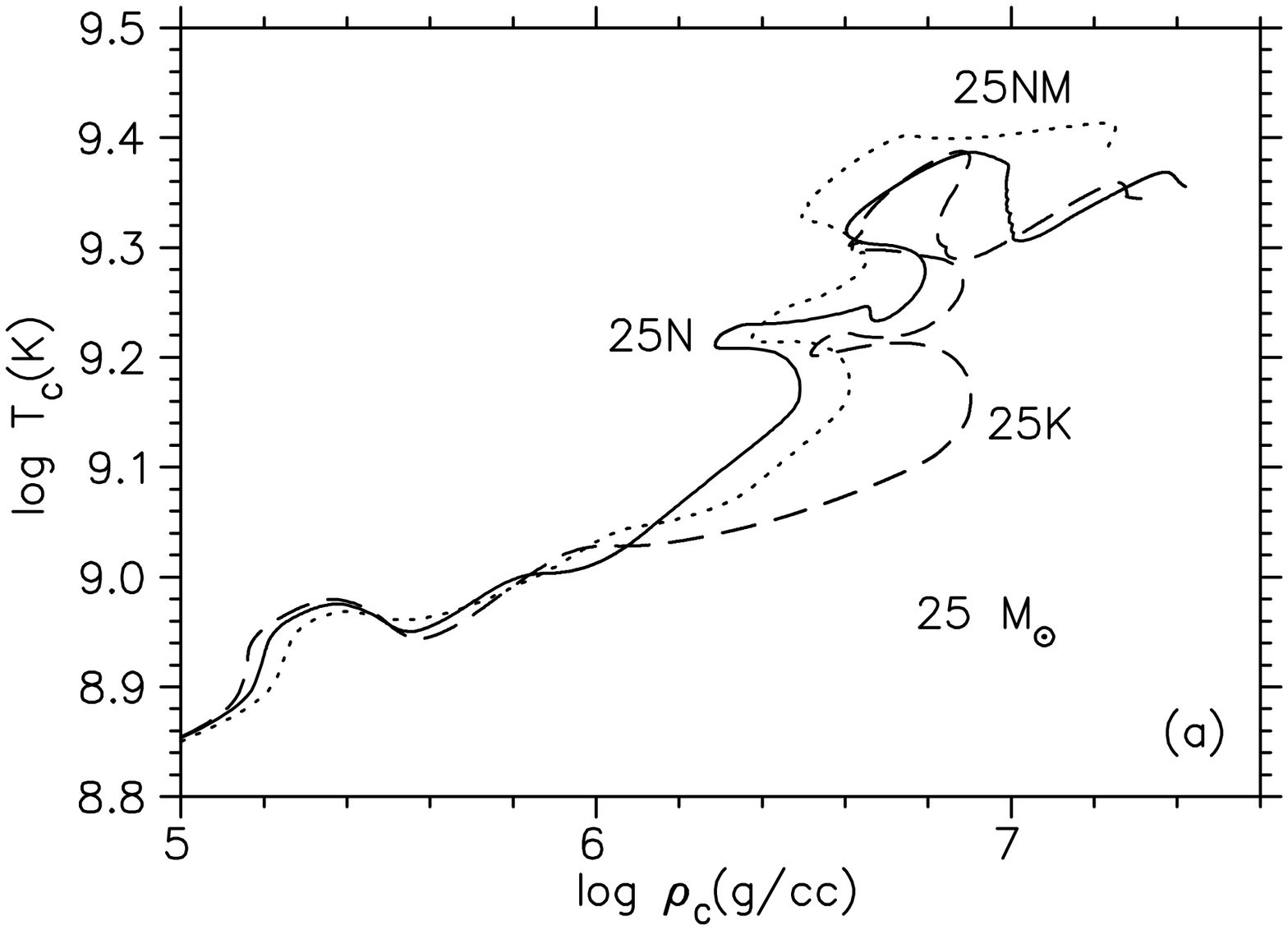}{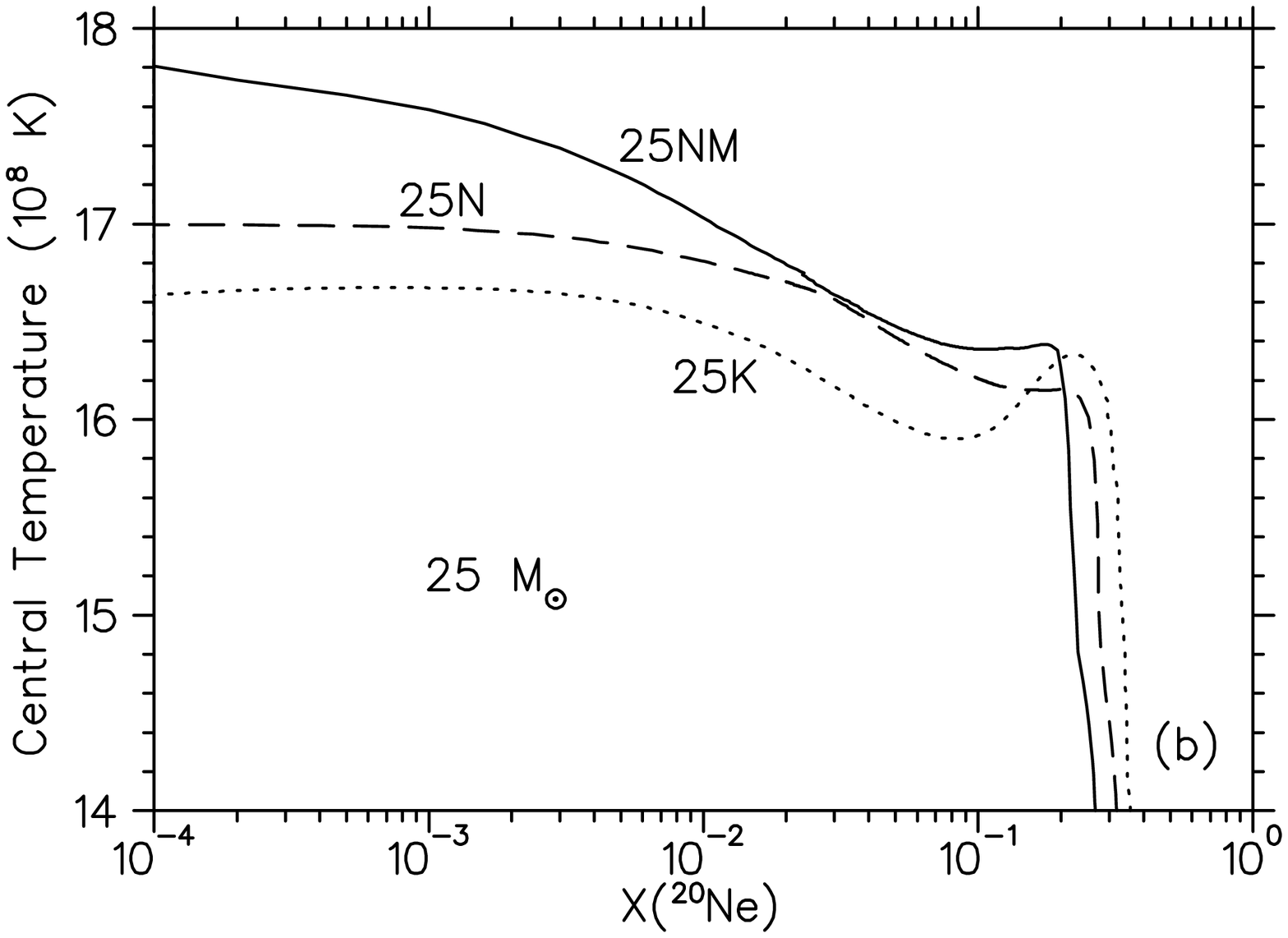}}
\caption{
(a) Similar to Fig. \ref{fig:Tc_vs_rhoC} showing only the evolution of the
center of the 25 $\msun$ star under different
assumptions (See Table \ref{tab:stellarModels}).
(b) Central Temperature T$_C$ versus the central
mass fraction of $^{20}$Ne, X($^{20}$Ne).
The curves illustrate how T$_C$ evolves differently
through the core Ne-burning phase under different
assumptions (see \ref{sec:neonphase} for details).
\label{fig:Tc_vs_rhoC_25msun}
}
\end{figure}
\clearpage

\begin{figure}
\figurenum{14}
\epsscale{0.7}
\rotatebox{90}{\plottwo{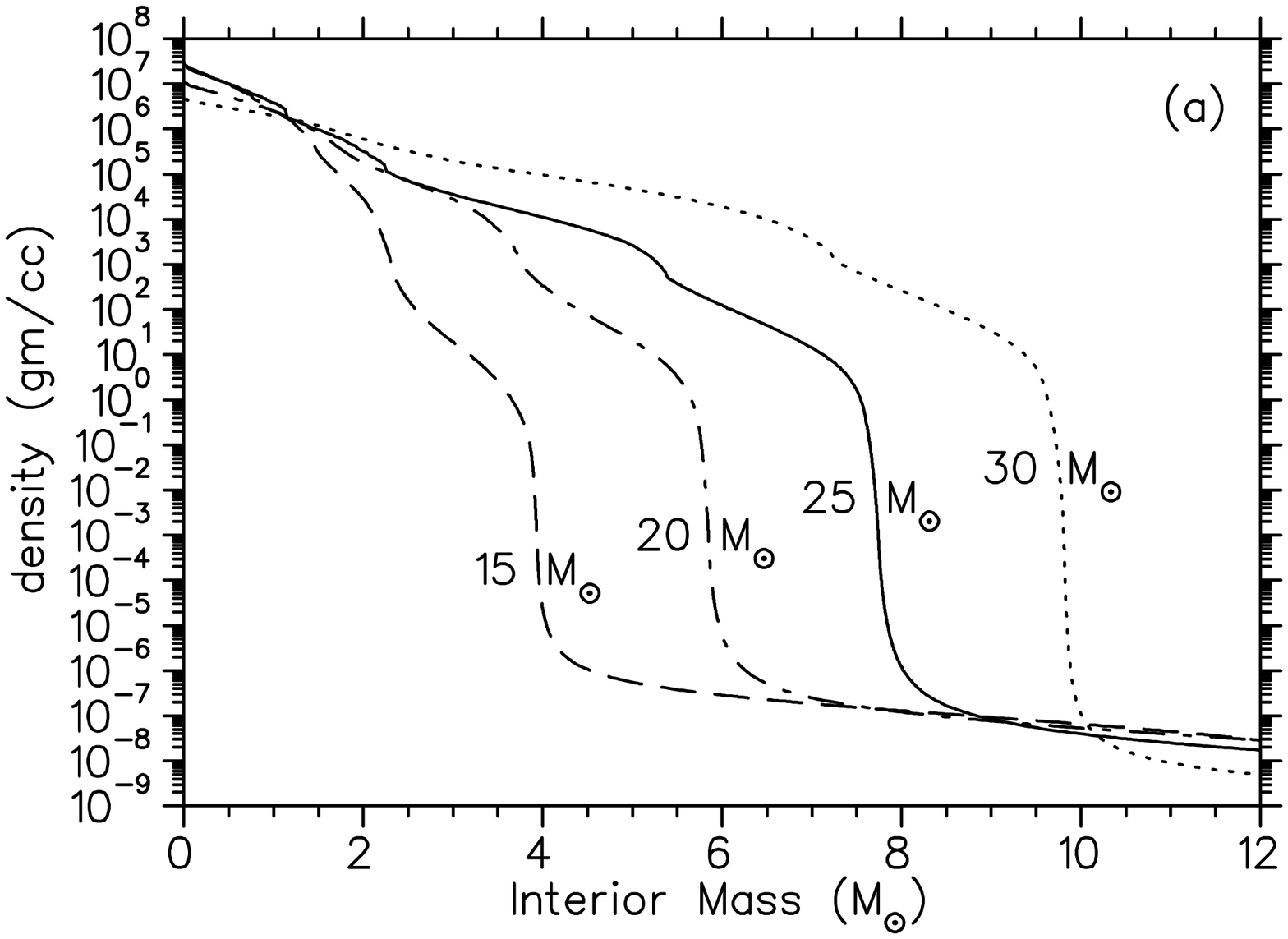}{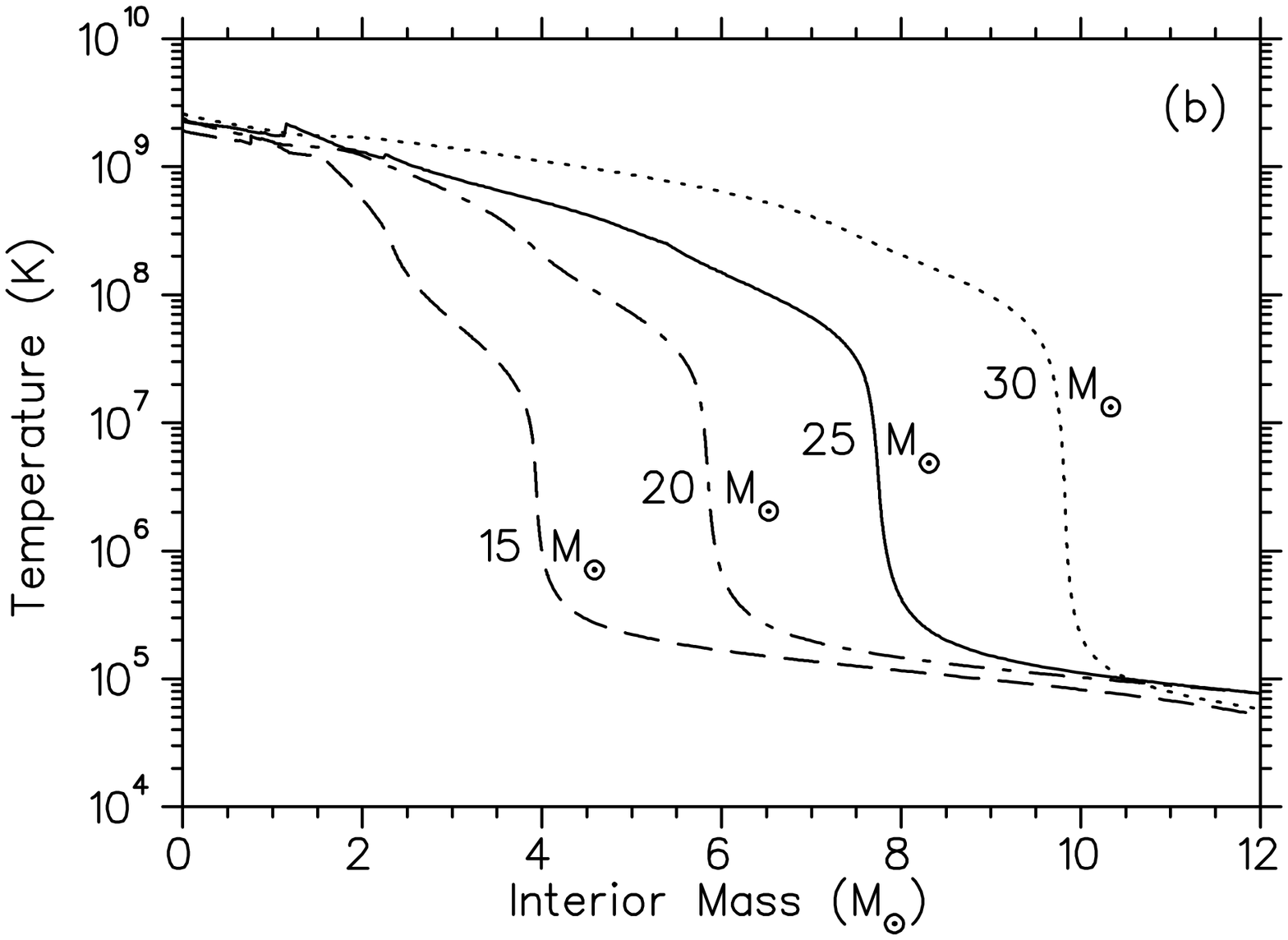}}
\caption{
Density (a) and temperature (b) versus interior mass 
for the stars of masses
15 $\msun$ to 30 $\msun$ as they evolved past core
oxygen burning.
\label{fig:dns_vs_Mr_at_end_Oburn}
}
\end{figure}
\clearpage


\begin{figure}
\figurenum{15}
\epsscale{0.7}
\rotatebox{90}{\plotone{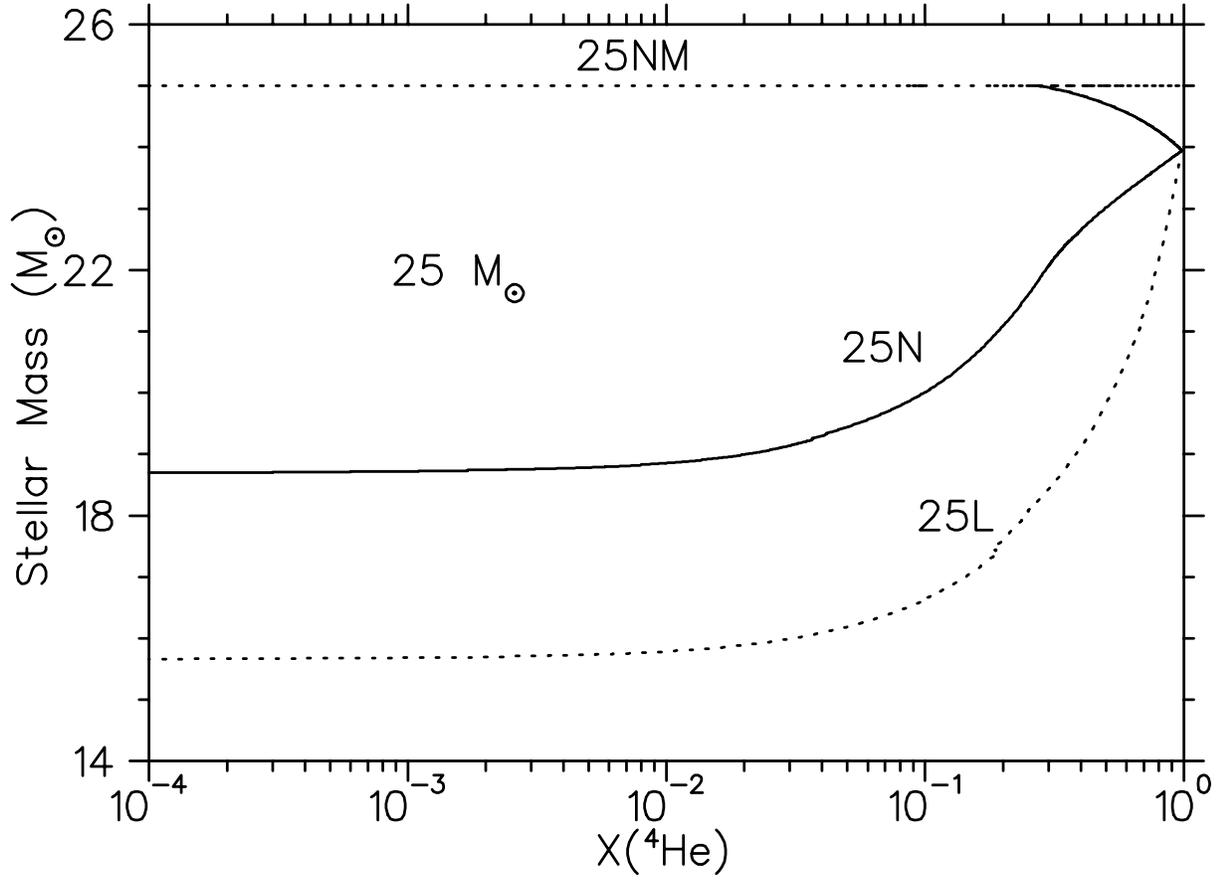}}
\caption{
Stellar mass versus central mass fraction of Helium,
X($^4$He) for the 25 $\msun$ star with and without
mass loss.
Notice the sharp decrease of the stellar mass during
the late phase of core He-burning (X($^4$He) $\leq$ 0.1).
\label{fig:Mass_vs_XHe4_25Msun_Heburn}
}
\end{figure}
\clearpage


\end{document}